%% file: root.tex
\documentclass{ifacconf}

\counterwithin*{section}{part}

\usepackage[english]{babel}
\usepackage{amssymb}

\usepackage{pgfplots}
\pgfplotsset{compat=newest}
\pgfplotsset{plot coordinates/math parser=false}
\newlength\figureheight
\newlength\figurewidth 

\usepackage{float}

\usepackage{tikz}
\usepackage{mathtools}
\usepackage{etoolbox}
\usepackage[ruled,vlined,linesnumbered,algo2e]{algorithm2e}
\makeatletter
\patchcmd{\algocf@Vline}{\vrule}{\vrule\vspace{-.32em}}{}{}
\makeatother
\SetStartEndCondition{ }{ }{}%
\SetKw{KwTo}{to}\SetKwFor{For}{for}{\string do}{}%
\SetKwIF{If}{ElseIf}{Else}{if}{then}{elif}{else}{}%
\SetKwFor{While}{while}{\string do}{}%

\DeclareMathAlphabet{\pazocal}{OMS}{zplm}{m}{n}
\usepackage [autostyle, english = american]{csquotes} 
\MakeOuterQuote{"}
\usepackage{xargs}
\usepackage{subfigure}

\usepackage{soul, color, xcolor}

\usepackage[numbers]{natbib}

\usepackage{cleveref}
\usepackage{empheq} 

\usetikzlibrary{positioning,arrows,petri,calc,decorations.markings,arrows.meta}
\tikzset{
place/.style={
circle,
thick,
minimum size=4mm,
draw
},
transitionV/.style={
rectangle,
thick,
fill=black,
minimum height=6mm,
inner xsep=1pt
}
}
\definecolor{myblue}{RGB}{0, 101, 202} 
\definecolor{mygreen}{RGB}{130, 180, 0} 
\definecolor{myred}{RGB}{197, 14, 31}

\input{math_commands.tex}

\begin{document}


\begin{frontmatter}

\title{Switched Max-Plus Linear-Dual Inequalities: 
 Application in Scheduling of Multi-Product Processing Networks\thanksref{footnoteinfo}} 

\author[1]{Davide Zorzenon}
\author[2]{Jan Komenda}
\author[1,3]{J\"{o}rg Raisch}

\thanks[footnoteinfo]{This work was funded by the Deutsche Forschungsgemeinschaft (DFG, German Research Foundation), Projektnummer RA 516/14-1. Partially supported by the GACR grant 19-06175J, by  MSMT INTER-EXCELLENCE project LTAUSA19098, by RVO 67985840, and by Deutsche Forschungsgemeinschaft (DFG, German Research Foundation) under Germany’s Excellence Strategy -- EXC 2002/1 "Science of Intelligence" -- project number 390523135.
}

\address[1]{Control Systems Group, Technische Universit\"at Berlin, Berlin, Germany (e-mail: [zorzenon,raisch]@control.tu-berlin.de)}
\address[2]{Institute of Mathematics, Czech Academy of Sciences, Prague, Czech Republic (e-mail: komenda@ipm.cz)}
\address[3]{Science of Intelligence, Research Cluster of Excellence, Berlin, Germany}

\begin{abstract}
P-time event graphs are discrete event systems suitable for modeling processes in which tasks must be executed in predefined time windows.
Their dynamics can be represented by systems of linear dynamical inequalities in the max-plus algebra and its dual, the min-plus algebra, referred to as max-plus linear-dual inequalities (LDIs).
We define a new class of models called switched LDIs (SLDIs), 
which allow to switch between different modes of operations, each corresponding to an LDI, 
according to an infinite sequence of modes called schedule.
In this paper, we focus on the analysis of SLDIs 
when the schedule is fixed and periodic.
We show that SLDIs 
can model single-robot multi-product processing networks, in which every product has different processing requirements and corresponds to a specific mode of operation.
Based on the analysis of SLDIs, 
we propose an algorithm to compute minimum and maximum cycle times for these processes that improves the time complexity of other existing approaches.
\end{abstract}


\end{frontmatter}

\section{Introduction}

P-time event graphs (P-TEGs) are event graphs in which tokens are forced to sojourn in places in predefined time windows.
They have been applied to solve scheduling problems for several processing networks
, including electroplating lines and cluster tools, cf.~\cite{becha2013modelling,kim2003scheduling}.
A common feature of these processing networks is that operations need to be executed in specified time intervals in order to obtain the desired quality of the final product, and P-TEGs are the ideal tools for modeling such constraints.

In this paper, we introduce a new class of systems called switched max-plus linear-dual inequalities (SLDIs). 
They extend the modeling power of P-TEGs by allowing to switch among different modes of operations, each consisting in a system of inequalities describing the dynamics of a P-TEG.
We first highlight the equivalence between bounded consistency, an important property extended to SLDIs from P-TEGs, and the existence of periodic trajectories.
SLDIs are then applied to model single-robot multi-product processing networks, namely, processing networks in which the type of products to be processed can change over time, each type requires to visit different processing stations, and products are transported by a single robot (see~\cite{KATS20081196} for a formal definition).
In this case, each mode is associated with a certain product type.

When the sequence of modes is fixed and periodic with period $|v|\in\nat$, the minimum and maximum cycle times of such systems can be computed in strongly polynomial time $\pazocal{O}(|v|^4n^4)$ (in the worst case) using an algorithm presented in~\cite{KATS20081196}, where $n$ corresponds to the total number of processing stations in the network.
We provide other two algorithms based on SLDIs that solve instances of the same problem.
The first one is derived from an existing procedure that computes the cycle times of P-TEGs, and achieves time complexity $\pazocal{O}(|v|^4n^4)$.
The second one, of time complexity $\pazocal{O}(|v|n^3+n^4)$, improves the first one by using tools from automata theory to exploit the sparsity of a certain matrix in the max-plus algebra. 
Tests are performed on an example of single-robot multi-product processing network to show the advantages of the proposed methods.

\subsection*{Notation}
The set of positive, respectively non-negative, integers is denoted by $\nat$, respectively $\nato$.
The set of non-negative real numbers is denoted by $\R_{\geq 0}$.
Moreover, $\Rmax \coloneqq \R \cup \{-\infty\}$, $\Rmin \coloneqq \R\cup\{\infty\}$, and $\Rbar \coloneqq \Rmax \cup \{\infty\}=\Rmin\cup\{-\infty\}$.
If $A\in\Rbar^{n\times n}$, we will use notation $A^\sharp$ to indicate $-A^\intercal$.

\section{Preliminaries}\label{se:preliminaries}

In the following subsections, some preliminary notions on idempotent semirings, precedence graphs, and multi-precedence graphs are recalled.
For a more detailed discussion on the first two topics, we refer to~\cite{baccelli1992synchronization} and~\cite{hardouin2018control}; multi-precedence graphs have been introduced in~\cite{zorzenon2021nonpositive}.

\subsection{Formal languages and the max-plus algebra}\label{su:idempotent_semirings}

A \textit{diod} (or \textit{idempotent semiring}) $(\D,\oplus,\otimes)$ is a set $\D$ endowed with two operations: $\oplus$ (addition), and $\otimes$ (multiplication).
Operation $\oplus$ and $\otimes$ are associative and have a neutral element indicated, respectively, by $\varepsilon$ and $e$; $\oplus$ is commutative and idempotent ($a \oplus a = a$), $\otimes$ distributes over $\oplus$, and $\varepsilon$ is absorbing for $\otimes$ ($\forall a\in\D$, $a\otimes \varepsilon = \varepsilon\otimes a = \varepsilon$).
For the sake of brevity, we will often omit symbol $\otimes$.
The order relation $\preceq$ is induced by $\oplus$ by: $a\preceq b\ \Leftrightarrow \ a\oplus b = a$.
A dioid is \textit{complete} if it is closed for infinite sums and $\otimes$ distributes over infinite sums.
In complete dioids, $\top=\bigoplus_{x\in\D}x$ denotes the greatest element of $\D$, the Kleene star of an element $a\in\D$ is defined by $a^* = \bigoplus_{k\in\nato}a^k$, where $a^0=e$, $a^{k+1} = a^k \otimes a$, and the \textit{dual addition} (or \textit{greatest lower bound}) $\splus$ is defined by $a\splus b = \bigoplus_{\D_{ab}}x$, where $\D_{ab}=\{x\in\D\ |\ x\preceq a \mbox{ and }x\preceq b\}$.

An example of a dioid that will be used in this paper is the algebra of formal languages.
Let $\Sigma=\{\wA_1,\ldots,\wA_l\}$ be a finite \textit{alphabet} of \textit{symbols} $\wA_1,\ldots,\wA_l$.
Then, $\Sigma^*$, respectively, $\Sigma^\omega$, indicate the set of all finite, respectively, infinite, sequences of symbols from $\Sigma$, called \textit{strings}.
Given two strings $s,t\in\Sigma^*$, their concatenation forms a new string $st\in\Sigma^*$; given a string $s\in\Sigma^*$ and a number $k\in\nato$, $s^k\in\Sigma^*$ and $s^\omega\in\Sigma^\omega$ denote, respectively, the string obtained by concatenating $s$ $k$, respectively, infinitely many times, with $s^0=\emptystr$, where $\emptystr$ denotes the empty string.
The length of a string $s$ is indicated by $|s|$ (with $|\emptystr|=0$), and $|s|_{\wA_i}$ is the number of occurrences of letter $\wA_i$ in $s$.
Moreover, $s_i$ indicates the $i$\textsuperscript{th} symbol of $s$, with $s_0=\emptystr$.
The prefix set of string $s\in\Sigma^*\cup\Sigma^\omega$ is defined by $\mbox{Pre}(s)=\{t_1\in\Sigma^*\ | \ t_1t_2 = s \mbox{ for some }t_2\in\Sigma^*\cup\Sigma^\omega\}$ and $s_{k]}$ indicates the string formed by the first $k$ symbols of $s$: $s_{k]} = s_1s_2\ldots s_k$, with $s_{0]} = \emptystr$.
We denote by $2^{\Sigma^*}$ the set of subsets of $\Sigma^*$.
Then, a (\textit{formal}) \textit{language} $\lang$ is an element of $2^{\Sigma^*}$, i.e., $\lang\subseteq\Sigma^*$.
The union of two languages $\lang_1,\lang_2\in 2^{\Sigma^*}$ is indicated by $\lang_1\cup\lang_2$, and $\lang_1\cdot\lang_2 = \lang_1\lang_2 = \{vw\ |\ v\in\lang_1,\ w\in\lang_2\}$ indicates the language obtained by concatenating all strings of $\lang_1$ with those of $\lang_2$.
Given a string $s\in\Sigma^*$, we will often indicate by the same symbol the single-string language $s\coloneqq \{s\}\in 2^{\Sigma^*}$.
It is easy to show that $(2^{\Sigma^*},\cup,\cdot)$ forms a complete dioid, in which $e = \{\emptystr\}$, $\varepsilon = \emptyset$, $\top = \Sigma^*$, $\splus$ coincides with $\cap$, and $\preceq$ coincides with $\subseteq$.

Before giving a second example of dioid -- the max-plus algebra -- we recall some other definitions and propositions.
Given $\MA,\MB\in\D^{m\times n}$, $\MC\in\D^{n\times p}$, operations $\oplus$ and $\otimes$ are extended to matrices as $(\MA\oplus\MB)_{ij}=\MA_{ij}\oplus\MB_{ij}$, and $(\MA\otimes\MC)_{ij} = \bigoplus_{k=1}^n (\MA_{ik}\otimes\MC_{kj})$.
Furthermore, the multiplication between a scalar $\lambda\in\D$ and a matrix $A\in\D^{m\times n}$ is defined by $(\lambda \otimes A)_{ij}=\lambda\otimes A_{ij}$.
If $(\D,\oplus,\otimes)$ is a complete dioid, then $(\D^{n\times n},\oplus,\otimes)$ is a complete dioid, too, with neutral elements for $\oplus$, $\otimes$, and $\splus$, respectively, given by the matrices $\pazocal{E}$, $E_\otimes$, and $\pazocal{T}$, where for all $i,j$, $\pazocal{E}_{ij}=\varepsilon$, $E_{\otimes ij} = e$ if $i=j$, $E_{\otimes ij}=\varepsilon$ else, and $\pazocal{T}_{ij}=\top$.
A binary operation $\stimes$ is called \textit{dual product} if it is associative, distributes over $\splus$, $e$ is its neutral element, and $\top$ is absorbing for $\stimes$.
Moreover, if $\stimes$ is a dual product for $(\D,\oplus,\otimes)$, then its extension to matrices, given by, $\forall A\in\D^{m\times n},C\in\D^{n\times p},\lambda\in\D$, $(A\stimes C)_{ij}=\bigsplus_{k=1}^n (A_{ik}\stimes C_{kj})$, $(\lambda\stimes A)_{ij}=\lambda\stimes A_{ij}$, is a dual product for $(\D^{n\times n},\oplus,\otimes)$.

Consider a complete \textit{idempotent semifield} $(\D,\oplus,\otimes)$, i.e., a complete dioid in which every element $a\in\D\setminus\{\varepsilon,\top\}$ admits a multiplicative inverse $a^{-1}$, i.e., $a\otimes a^{-1}=a^{-1}\otimes a = e$.
Then, operation $\stimes$ defined as $a\stimes b = a\otimes b$ if $a,b\in\D\setminus\{\top\}$, $a\stimes b=\top$ if $a=\top$ or $b=\top$ is a dual product for $(\D,\oplus,\otimes)$ (see~\cite{zorzenon2021periodic}).

The \textit{tensor} (or \textit{Kronecker}) \textit{product} $\otimes^t$ between two matrices $A\in\D^{m\times n}$, $B\in\D^{p\times q}$ is defined as the matrix 
\[
	A\otimes^t B = \begin{bmatrix}
		A_{11} \otimes B & \cdots & A_{1n}\otimes B\\
		\vdots & & \vdots\\
		A_{m1} \otimes B & \cdots & A_{mn} \otimes B
	\end{bmatrix}
	\in\D^{mp\times nq}.
\]
We recall the following properties of $\otimes^t$, the first of which holds in \textit{commutative dioids}, i.e., dioids in which $\otimes$ is commutative.

\begin{prop}[\cite{horn1991topics}]\label{pr:tensor_mixed}
Let $(\D,\oplus,\otimes)$ be a commutative diod, $A\in\D^{m\times n}$, $B\in\D^{p\times q}$, $C\in\D^{n\times k}$, $D\in\D^{q\times r}$.
Then $(A\otimes^t B)\otimes (C\otimes^t D) = (A\otimes C)\otimes^t (B\otimes D)$.
\end{prop}

\begin{prop}\label{pr:trace_tensor}
Let $(\D,\oplus,\otimes)$ be a dioid, $A\in\D^{m\times m}$, $B\in\D^{p\times p}$.
Then, $\tr(A\otimes^t B) = \tr(A)\otimes \tr(B)$, where $\tr(M) = \bigoplus_{i = 1}^q M_{ii}$ indicates the trace of matrix $M\in\D^{q\times q}$.
\end{prop}
\begin{pf}
	\[
		\begin{array}{rl}
		\tr(A\otimes^t B) =& \bigoplus_{k = 1}^m \tr(A_{kk}\otimes B) = \bigoplus_{k = 1}^m A_{kk}\otimes \tr(B) \\ =& \left(\bigoplus_{k = 1}^m A_{kk}\right) \otimes \tr(B) = \tr(A)\otimes \tr(B).
	\quad\qed
		\end{array}
	\]
\end{pf}

The \textit{max-plus algebra} is the complete and commutative idempotent semifield $(\Rbar,\oplus,\otimes)$, i.e., the set of extended real numbers endowed with the standard maximum operation $\oplus$, and the standard addition $\otimes$.
In the max-plus algebra, $\varepsilon=-\infty$, $e = 0$, $\top = \infty$, $\splus$ is the standard minimum operation, $\preceq$ coincides with $\leq$.
The dual product $\stimes$ is such that $a\stimes b= a\otimes b$ if $a,b\neq \infty$, and $a\stimes b = \infty$ if $a = \infty$ or $b = \infty$.
The extension of the max-plus algebra to square matrices $(\Rbar^{n\times n},\oplus,\otimes)$ is a complete dioid; in the rest of the paper, symbol $\preceq$ will be reserved to compare matrices with elements from $\Rbar$, i.e., $\forall A,B\in\Rbar^{m\times n}$, $A\preceq B \Leftrightarrow A_{ij}\leq B_{ij}$ $\forall i,j$.
The product between a scalar $\lambda\in\R$ and a matrix $A\in\Rbar^{n\times n}$, $\lambda\otimes A = \lambda\stimes A$, will simply be indicated by $\lambda A$.
Note that, with the notation above, $(\Rbar,\splus,\stimes)$ forms a dual dioid called the \textit{min-plus algebra}.

\subsection{Precedence graphs and multi-precedence graphs}\label{su:precedence_graphs}

A \textit{directed graph} is a pair $(\nodes,\arcs)$ where $\nodes$ is a finite set of nodes and $\arcs\subseteq\nodes\times\nodes$ is the set of arcs.
A \textit{weighted directed graph} is a triplet $(\nodes,\arcs,w)$, where $(\nodes,\arcs)$ is a directed graph, and $w:\arcs\rightarrow\R$ is a function that associates a weight $w((i,j))$ to each arc $(i,j)\in\arcs$ of graph $(\nodes,\arcs)$.

The \textit{precedence graph} associated with a matrix $\MA\in\Rmax^{n\times n}$ is the weighted directed graph $\graph(A)=(\nodes,\arcs,w)$, where $\nodes=\{1,\ldots,n\}$, and $\arcs$ and $w$ are defined in the following (non-standard) way: there is an arc $(i,j)\in\arcs$ from node $i$ to node $j$ if and only if $\MA_{ij}\neq-\infty$, and $w$ is such that $w((i,j))=A_{ij}$.
We adopt this non-standard convention of associating $A_{ij}$ to the weight of arc $(i,j)$ instead of $(j,i)$, as this will simplify the interpretation of the label of a path in multi-precedence graphs.
When elements of $A$ are functions of some real parameters, $\MA=\MA(\lambda_1,\ldots,\lambda_p)$, $\lambda_1,\ldots,\lambda_p\in\R$, we say that $\graph(A)$ is a \textit{parametric precedence graph}.
A sequence of $r+1$ nodes $\rho=(i_1,i_2,\ldots,i_{r+1})$, $r\geq 1$, such that $(i_j,i_{j+1})\in\arcs$ for all $j\in\{1,\ldots,r\}$ is a path of length $r$; a path $\rho$ such that $i_1 = i_{r+1}$ is called a circuit.
The weight of a path is the sum (in conventional algebra) of the weights of the arcs composing it.
Elements of the max-plus power of a matrix $A$ have a clear meaning with respect to precedence graph $\graph(A)$; indeed, $(A^r)_{ij}$ corresponds to the maximum weight of all paths in $\graph(A)$ of length $r$ from node $i$ to node $j$.
The maximum circuit mean of a precedence graph $\graph(A)$ with $n$ nodes can be computed in the max-plus algebra as $\mbox{mcm}(A)=\bigoplus_{k=1}^n \tr(A^k)^{\frac{1}{k}}$, where $a^{\frac{1}{k}}$ is the $k$\textsuperscript{th} max-plus root of $a\in\Rmax$ and corresponds to $\frac{a}{k}$ in standard algebra. 
We recall that a precedence graph $\graph(A)$ does not contain circuits with positive weight if and only if $\tr(A^*) = 0$; otherwise, if there is at least one circuit with positive weight in $\graph(A)$, then $\tr(A^*) = \infty$.

In this paper, we will make use of another class of graphs, called \textit{multi-precedence graphs}, which will allow us to analyze parametric precedence graphs using tools from formal languages and automata theory.
The reader familiar with max-plus automata will notice their similarity to multi-precedence graphs.
The multi-precedence graph associated with matrices $\MA_1,\ldots,\MA_l\in\Rmax^{n\times n}$ is the weighted multi-directed graph $\graph(A_1,\ldots,A_l) = (\nodes,\Sigma,\ltm,\arcs)$, where $\nodes=\{1,\ldots,n\}$ is the set of nodes, $\Sigma=\{\wA_1,\ldots,\wA_l\}$ is the alphabet of symbols $\wA_1,\ldots,\wA_l$, $\ltm:\Sigma\rightarrow\Rmax^{n\times n}$ is the morphism defined by $\ltm(\wA_i) = \MA_i$ for all $i\in\{1,\ldots,l\}$, and $\arcs\subseteq \nodes\times\Sigma\times\nodes$ is the set of labeled arcs, defined such that there is an arc $(i,\wZ,j)\in\arcs$ from node $i$ to node $j$ labeled $\wZ$ with weight $(\ltm(\wZ))_{ij}$ if and only if $(\ltm(\wZ))_{ij}\neq -\infty$.
A path in a multi-precedence graph $\graph(\MA_1,\ldots,\MA_l)$ is a sequence of alternating nodes and labels of the form $\sigma=(i_1,\wZ_1,i_2,\wZ_2,\ldots,\wZ_r,i_{r+1})$, $r\geq 1$, such that $(i_j,\wZ_j,i_{j+1})\in\arcs$ for all $j=1,\ldots,r$; we will say that path $\sigma$ is labeled $s=\wZ_1\wZ_{2}\ldots\wZ_r$.

It is convenient to extend morphism $\ltm$ to $\ltm:2^{\Sigma^*}\rightarrow\Rbar^{n\times n}$ as follows:
for all $\wZ\in\Sigma$, $\lang_1,\lang_2\subseteq\Sigma^*$,	$\ltm(\{\emptystr\}) = E_\otimes$, $\ltm(\{\wZ\}) = \ltm(\wZ)$, $\ltm(\lang_1\cup\lang_2) = \ltm(\lang_1)\oplus\ltm(\lang_2)$, and $\ltm(\lang_1\lang_2) = \ltm(\lang_1)\otimes\ltm(\lang_2)$. 
In this way, given a language $\lang\subseteq\Sigma^*$, $\ltm(\lang)_{ij} = \bigoplus_{s\in\lang}\ltm(s)_{ij}$ corresponds to the supremum, for all strings $s\in\lang$, of the weights of all paths labeled $s$ in $\graph(\MA_1,\ldots,\MA_l)$ from node $i$ to node $j$; in particular, $\tr(\ltm(\lang)^*)=0$ if and only if no circuits with positive weight exist in $\graph(A_1,\ldots,A_l)$ among those with label $s\in\lang$.
Moreover, the following properties hold: for all $\lang_1,\lang_2,\lang\subseteq\Sigma^*$, $\lang_1\subseteq\lang_2 \Rightarrow \ltm(\lang_1)\preceq\ltm(\lang_2)$, and $\ltm(\lang^*) = \ltm(\lang)^*$.
We will indicate by $\nonegset$, respectively, $\nonegsetm$, the set of all precedence graphs, respectively, multi-precedence graphs, that do not contain circuits with positive weight.
The following proposition allows us to study the sign of circuit weights in some precedence graphs using multi-precedence graphs.

\begin{prop}[\cite{zorzenon2021nonpositive}]\label{pr:multi_precedence}
	Let $\MA_1,\ldots,\MA_l\in\Rmax^{n\times n}$.
	There exists a circuit with positive weight visiting node $i\in\{1,\ldots,n\}$ in multi-precedence graph $\graph(\MA_1,\ldots,\MA_l)$ if and only if there exists a circuit with positive weight visiting node $i$ in precedence graph $\graph(\MA_1\oplus\ldots\oplus\MA_l)$.
\end{prop}

Given a parametric precedence graph $\graph(\MA)$, where $\MA = \MA(\lambda_1,\ldots,\lambda_l)$, the \textit{non-positive circuit weight problem} (NCP) consists in characterizing the set $\solNCP{\MA} = \{(\lambda_1,\dots,\lambda_l)\in\R^l\ |\ \graph(\MA)\in\nonegset\}$ of all values of parameter $(\lambda_1, \dots,\lambda_l)$ for which $\graph(A)$ does not contain circuits with positive weight.
When matrix $\MA$ has the form $\MA(\lambda) = \lambda\MP \oplus \lambda^{-1}\MI \oplus \MC$ for arbitrary matrices $\MP,\MI,\MC\in\Rmax^{n\times n}$ (called proportional, inverse, and constant matrix, respectively), then $\solNCP{\lambda\MP\oplus\lambda^{-1}\MI\oplus\MC}=[\lambda_{\text{min}},\lambda_{\text{max}}]\cap\R$ is an interval; moreover, its extremes can be found either in weakly polynomial time using linear programming solvers such as the interior-point method, or in strongly polynomial time $\pazocal{O}(n^4)$ using Algorithm~\ref{al:PIC-NCP}, see~\cite{zorzenon2021nonpositive}.

\begin{algorithm2e}[t] \DontPrintSemicolon \label{al:PIC-NCP} \small
\KwIn{$\MP,\MI,\MC\in \Rmax^{n\times n}$}
\KwOut{$\Lambda_{\text{NCP}}(\lambda\MP\oplus \lambda^{-1}\MI\oplus \MC)$}
 \lIf{$\graph(\MC)\notin \nonegset$}{%
     \Return $\emptyset$%
 }%
$\MP \leftarrow \MC^* \MP \MC^*$, $\MI \leftarrow \MC^* \MI \MC^*$, $S \leftarrow E_{\otimes}$\\
\For{$k = 1$ \KwTo $\left \lfloor{\frac{n}{2}} \right \rfloor$}{\vspace*{2pt}
	$S\leftarrow \MP S^2 \MI \oplus \MI S^2 \MP \oplus E_{\otimes}$
}
\lIf{$\graph(S)\notin \nonegset$}{%
	\Return $\emptyset$%
}%
\Return $[\mbox{mcm}(\MI S^*),(\mbox{mcm}(\MP S^*))^{-1}]\cap\R$
 \caption{$\mathsf{Solve\_NCP}(\MP,\MI,\MC)$}
\end{algorithm2e}

\section{P-time event graphs}\label{se:P-time_event_graphs}

\begin{defn}[From~\cite{CALVEZ19971487}]
	An unweighted P-time Petri net (P-TPN) is a 5-tuple $(\places,\transitions,\arcs,m,\iota)$, where $(\places\cup\transitions,E)$ is a directed graph in which the set of nodes is partitioned into the set of places, $\places$, and the set of transitions, $\transitions$, the set of arcs $\arcs$ is such that $\arcs\subseteq (\places\times \transitions)\cup(\transitions\times\places)$, $m:\places\rightarrow\nato$ is a map such that $m(p)$ represents the number of tokens initially residing in place $p\in\places$ (also called initial marking of $p$), and $\iota:\places\rightarrow\{[\tau^-,\tau^+]\ |\ \tau^-\in\R_{\geq 0},\tau^+\in\R_{\geq 0}\cup\{\infty\},\tau^-\leq \tau^+\}$
	is a map that associates to every place $p\in\places$ a time interval $\iota(p)=[\tau^-_p,\tau^+_p]$.
\end{defn}

The dynamics of a P-TPN net is briefly described as follows.
A transition $t$ is enabled when either it has no upstream place or each upstream place $p$ of $t$ contains at least one token which has resided in $p$ for a time between $\tau^-_{p}$ and $\tau^+_{p}$ (extremes included).
When transition $t$ is enabled, it may fire; its firing causes one token to be removed instantaneously from each of the upstream places of $t$, and one token to be added, again instantaneously, to each of the downstream places of $t$.
If a token sojourns more than $\tau^+_{p}$ time instants in a place $p$, then said token is \textit{dead}, as it is forced to remain in $p$ forever.

A P-time event graph (P-TEG) is a P-TPN in which every place has exactly one upstream and one downstream transition.
Without loss of generality (see~\cite{vspavcek2017analysis}), we will suppose that the initial marking $m(p)$ is less than or equal to $1$ for each place $p\in\places$ of a P-TEG.
This allows to rephrase the dynamics of a P-TEG with $|\transitions| = n$ transitions as a max-plus linear-dual inequality system (LDI), i.e., a system of dynamical $(\oplus,\otimes)$- and $(\splus,\stimes)$-linear inequalities of the form
\begin{equation}\label{eq:dynamics_PTEGs}
\forall k\in\nato, \quad 
	\left\{
	\begin{array}{rcl}
		A^0\otimes x(k) \preceq & x(k) & \preceq B^0\stimes x(k)\\
		A^1\otimes x(k) \preceq & x(k+1) & \preceq B^1\stimes x(k)
	\end{array}
	\right.
	~,
\end{equation}
where $x:\nato\rightarrow \R^n$ is called \textit{dater function}, $A^0,A^1\in\Rmax^{n\times n}$, $B^0,B^1\in\Rmin^{n\times n}$ are called \textit{characteristic matrices} of the P-TEG, and are defined as follows.
If there exists a place $p$ with initial marking $\mu\in\{0,1\}$, upstream transition $t_j$ and downstream transition $t_i$, then $A^\mu_{ij}=\tau^-_p$ and $B^\mu_{ij}=\tau^+_p$; otherwise, $A^\mu_{ij} = -\infty$ and $B^\mu_{ij} = \infty$.
By convention, element $x_i(k)$ of the dater function represents the time at which transition $t_i$ fires for the $(k+1)$\textsuperscript{st} time.
Since the $(k+2)$\textsuperscript{nd} firing of any transition cannot occur before the $(k+1)$\textsuperscript{st}, we require the dater to be a non-decreasing function, i.e., $\forall i\in\{1,\ldots,n\}$, $x_i(k+1)\geq x_i(k)$.

If a non-decreasing dater trajectory $\{x(k)\}_{k\in\nato}$ satisfying~\eqref{eq:dynamics_PTEGs} exists, then the trajectory is said to be \textit{consistent} for the P-TEG, as it does not cause the death of any token, and the P-TEG is said to be \textit{consistent}.
A trajectory $\{x(k)\}_{k\in\nato}$ is 1-periodic with period $\lambda\in\R_{\geq 0}$, if it has the form $\{\lambda^kx(0)\}_{k\in\nato}$, in the max-plus algebra sense; in standard algebra, this corresponds to a dater trajectory such that, for all $i\in\{1,\ldots,n\}$, $x_i(k) = k \lambda + x_i(0)$.
Moreover, we indicate by $\solPTEG\subseteq\R$ the set of $\lambda\geq 0$ for which there exists a 1-periodic trajectory of period $\lambda$ that is consistent for the P-TEG characterized by matrices $A^0,A^1,B^0,B^1$; such periods are called cycle times.
We say that a trajectory $\{x(k)\}_{k\in\nato}$ is \textit{delay-bounded} if there exists a positive real number $M$ such that, for all $i,j\in\{1,\ldots,n\}$ and for all $k\in\nato$, $x_i(k)-x_j(k)<M$; a P-TEG admitting a consistent delay-bounded trajectory of the dater function is said to be \textit{boundedly consistent}.
To our knowledge, no algorithm that checks whether a P-TEG is consistent has been found until now;
on the other hand, there exists an algorithm that checks bounded consistency of P-TEGs in time $\pazocal{O}(n^4)$, which comes directly from the following result.

\begin{thm}[\cite{zorzenon2020bounded,zorzenon2021periodic}]\label{th:bounded_consistency}
	A P-TEG is\linebreak boundedly consistent if and only if it admits a consistent 1-periodic trajectory, i.e., if and only if set $\solPTEG$ is non-empty.
	Moreover, \linebreak$\solPTEG$ coincides with 
	\[
		\solNCP{\lambda B^{1\sharp}\oplus \lambda^{-1}A^1\oplus (A^0\oplus B^{0\sharp})} \cap [0,\infty[ .
	\]
\end{thm}


\begin{figure}
	\centering
	\resizebox{.5\linewidth}{!}{
		\input{figures/P-TEG_example.tex}
	}
	\caption{Example of P-TEG.}
	\label{fig:P-TEG_example}
\end{figure}
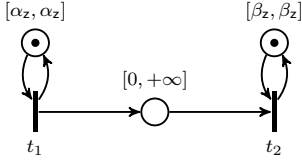
\begin{table}[t]
	\begin{center}
	\caption{Parameters for the P-TEG of~\Cref{fig:P-TEG_example}.} \label{tab:P-TEG_parameters}
		\begin{tabular}{ccc}
			$\wZ$ & $\alpha_\wZ$ & $\beta_\wZ$ \\\hline
			$\wA$ & 2 & 1 \\
			$\wB$ & 1 & 2 \\ 
			$\wC$ & 1 & 1
		\end{tabular}
	\end{center}
\end{table}

\begin{exmp}\label{ex:P-TEGs}
	Consider the P-TEG represented in~\Cref{fig:P-TEG_example}, in which time windows are parametrized with respect to label $\wZ$; in~\Cref{tab:P-TEG_parameters}, values of time windows are given for $\wZ\in\{\wA,\wB,\wC\}$.
	The matrices characterizing the P-TEG labeled $\wZ$ are:
	\[
		A^0_\wZ = 
		\begin{bmatrix}
			-\infty & -\infty \\
			0 & -\infty
		\end{bmatrix},\quad
		A^1_\wZ = 
		\begin{bmatrix}
			\alpha_\wZ & -\infty \\
			-\infty & \beta_\wZ
		\end{bmatrix},
	\]\[
		B^0_\wZ = 
		\begin{bmatrix}
			\infty & \infty \\
			\infty & \infty
		\end{bmatrix},\quad
		B^1_\wZ = 
		\begin{bmatrix}
			\alpha_\wZ & \infty \\
			\infty & \beta_\wZ
		\end{bmatrix}.
	\]
	Since lower and upper bounds for the sojourn times of the two places with an initial token coincide, once dater $x_\wZ(0)$ is chosen (such that the first inequality in~\eqref{eq:dynamics_PTEGs} is satisfied for $k=0$, i.e., $x_{\wZ,2}(0)\geq x_{\wZ,1}(0)$), the only trajectory $\{x_\wZ(k)\}_{k\in\nato}$ that is a candidate to be consistent for the P-TEG labeled $\wZ$ is deterministically given by
	\[
		\forall k\in\nato,\quad x_\wZ(k+1) = \begin{bmatrix}\alpha_\wZ + x_{\wZ,1}(k)\\\beta_\wZ + x_{\wZ,2}(k)\end{bmatrix}.
	\]
	However, it is easy to see that, for any valid choice of the initial dater, candidate trajectory $\{x_\wA(k)\}_{k\in\nato}$ is not consistent (as for a sufficiently large $k$, $x_{\wA,2}(k) < x_{\wA,1}(k)$
	), and $\{x_\wB(k)\}_{k\in\nato}$, despite being consistent, is not delay-bounded and results in the infinite accumulation of tokens in the place between $t_1$ and $t_2$ for $k\rightarrow\infty$.
	On the other hand, $\{x_\wC(k)\}_{k\in\nato}$ is consistent and delay-bounded (in fact, it is 1-periodic with period $1$); thus we can conclude that the P-TEG labeled $\wA$ is not consistent, the one labeled $\wB$ is consistent but not boundedly consistent, and the one labeled $\wC$ is boundedly consistent.
	Of course, we would have reached the same conclusions regarding delay-boundedness by using Theorem~\ref{th:bounded_consistency}.
	In particular, applying Algorithm~\ref{al:PIC-NCP}, we get
	\[
		\solPTEG[A^0_\wA][A^1_\wA][B^0_\wA][B^1_\wA] = \solPTEG[A^0_\wB][A^1_\wB][B^0_\wB][B^1_\wB] = \emptyset,
	\]\[
		\solPTEG[A^0_\wC][A^1_\wC][B^0_\wC][B^1_\wC] = [1,1] = \{1\}.
	\]
\end{exmp}

\section{Switched max-plus linear-dual inequalities}\label{se:switched_max-plus_linear-dual_systems}

\subsection{General description}\label{su:general_description}

We start by defining a switched LDI (SLDI) as the natural extension of the dynamical inequalities of P-TEGs, in which the mode of operation can switch.
Each mode is associated with a set of $n$ events that have to satisfy certain time window constraints.
An SLDI is a 5-tuple $\pazocal{S}=(\Sigma,A^0,A^1,B^0,B^1)$, where $\Sigma=\{\wA_1,\ldots,\wA_m\}$ is a finite alphabet whose symbols are called \textit{modes}, and $A^0,A^1:\Sigma \rightarrow \Rmax^{n\times n}$, $B^0,B^1:\Sigma\rightarrow \Rmin^{n\times n}$ are functions that associate a matrix to each mode of $\Sigma$; for sake of simplicity, given a mode $\wZ\in\Sigma$, we will write $A^0_\wZ,A^1_\wZ,B^0_\wZ,B^1_\wZ$ in place of $A^0(\wZ),A^1(\wZ),B^0(\wZ),B^1(\wZ)$, respectively.
A \textit{schedule} $w\in\Sigma^\omega$ is an infinite concatenation of modes.

The dynamics of an SLDI $\pazocal{S}$ under schedule $w\in\Sigma^\omega$ is expressed by the following system of inequalities: for all $k\in\nato$,
\begin{equation}\label{eq:dynamics}
	\left\{
	\begin{array}{rcl}
		A^0_{w_{k+1}}\otimes x(w_{k]}) \preceq & x(w_{k]}) & \preceq B^0_{w_{k+1}}\stimes x(w_{k]})\\
		A^1_{w_{k+1}}\otimes x(w_{k]}) \preceq & x(w_{k+1]}) & \preceq B^1_{w_{k+1}}\stimes x(w_{k]})
	\end{array}
	\right.
	~,
\end{equation}
where function $x:\mbox{Pre}(w)\rightarrow \R^n$ is called \textit{dater} of $\pazocal{S}$ associated with schedule $w$.
Term $x_i(w_{k]})$ represents the time of the occurrence of event $i$ associated with mode $w_{k+1}$.

When schedule $w$ is fixed, we can extend the definition of some properties of P-TEGs to SLDIs in a natural way.
For instance, if there exists a trajectory of the dater $\{x(w_{k]})\}_{k\in\nato}$ that satisfies~\eqref{eq:dynamics} for all $k\in\nato$, then the trajectory is consistent for the SLDI under schedule $w$, and we say that the SLDI is consistent under schedule $w$.
The definitions of delay-bounded trajectory and bounded consistency are generalized to SLDIs under schedule $w$ in a similar way.

The interpretation of bounded consistency of an SLDI under a fixed schedule $w$ is analogous to the one of P-TEGs (see~\cite{zorzenon2020bounded}).
When a process consisting of several tasks (each represented by an event) is modeled by an SLDI that is not boundedly consistent under a schedule $w$, then the execution of every possible sequence of tasks following $w$ 
will either lead to the violation of some time window constraints (if the SLDI is not consistent under $w$), or to the infinite accumulation of delay between the execution of some tasks (if the only consistent trajectories are not delay-bounded).

\subsection{Analysis of fixed periodic schedules}\label{su:analysis_of_fixed_schedules}

In this subsection, we analyze bounded consistency and cycle times of an SLDI when schedule $w$ is periodic, i.e., when it can be written as $w = v^\omega$, where $v\in\Sigma^*$ is a finite \textit{subschedule}.
Similarly to P-TEGs, it is natural to assume the following non-decreasingness condition for the dater of an SLDI: for all $k\in\nato$, $h\in\{0,\ldots,|v|-1\}$, $x(v^{k+1}v_1v_2\ldots v_h) \succeq x(v^kv_1 v_2 \ldots v_h)$.
The meaning is that events occurring during the $(k+2)$\textsuperscript{nd} repetition of mode $v_h$, at the $h$\textsuperscript{th} position in subschedule $v$, cannot occur earlier than those taking place during the $(k+1)$\textsuperscript{st} one.

We define $v$-periodic trajectories of period $\lambda\in\R_{\geq 0}$ for SLDIs under schedule $w=v^\omega$ as those dater trajectories that, for all $k\in\nato$, $h\in\{0,\ldots,|v|-1\}$, satisfy $x(v^{k+1}v_1\ldots v_h) = \lambda x(v^k v_1 \ldots v_h)$; $\solSLDI$ denotes the set of all periods $\lambda$, called cycle times, for which there exists a consistent $v$-periodic trajectory.
Their relationship with 1-periodic trajectories in P-TEGs is made clear by the following example.

\begin{exmp}\label{ex:2}
	Let us analyze the SLDI $\pazocal{S}$, with $\Sigma=\{\wA,\wB,\wC\}$, and $A^0_\wZ,A^1_\wZ,B^0_\wZ,B^1_\wZ$ defined as in Example~\ref{ex:P-TEGs}; now label $\wZ\in\Sigma$ is to be interpreted as a mode. 
	Thus, for each event $k$, the dynamics of the SLDI may switch among those specified by the P-TEGs labeled $\wA$, $\wB$, and $\wC$.
	We consider periodic schedules $(\wA\wC)^\omega$ and $(\wA\wB)^\omega$; observe that for $w=v^\omega$, with $v\in\{\wA\wC,\wA\wB\}$ (i.e., $v_1=\wA$ and $v_2=\wC$ or $v_2=\wB$) the SLDI following $w$ can be written as: for all $k\in\nato$,
	\begin{equation}\label{eq:example_1}
		\left\{
		\begin{array}{rcl}
			A^0_{v_1}\otimes x(v^k) \preceq & x(v^k) & \preceq B^0_{v_1}\stimes x(v^k)\\
			A^1_{v_1}\otimes x(v^k) \preceq & x(v^kv_1) & \preceq B^1_{v_1}\stimes x(v^k)\\
			A^0_{v_2}\otimes x(v^kv_1) \preceq & x(v^kv_1) & \preceq B^0_{v_2}\stimes x(v^kv_1)\\
			A^1_{v_2}\otimes x(v^kv_1) \preceq & x(v^{k+1}) & \preceq B^1_{v_2}\stimes x(v^kv_1)
		\end{array}
		\right.~.
	\end{equation}
	By defining $\tilde{x}(k) = [x^\intercal(v^k),x^\intercal(v^kv_1)]^\intercal$, the above set of inequalities can be rewritten as an LDI: for all $k\in\nato$, 
	\begin{subequations}\label{eq:example_2}
		\begin{empheq}[left=\empheqlbrace,right={~,}]{align}
			&A^0_{v}
	\otimes \tilde{x}(k) \preceq \,\,\,\,\, \tilde{x}(k) \,\,\,\,\, \preceq B^0_v \stimes \tilde{x}(k) \label{eq:example_2-1}\\
			&A^1_v \otimes \tilde{x}(k) \preceq  \tilde{x}(k+1) \preceq B^1_v \stimes \tilde{x}(k)\label{eq:example_2-2}
		\end{empheq}
	\end{subequations}
	where
	\[
		A^0_v = \begin{bmatrix}
			A^0_{v_1} & \pazocal{E} \\
			A^1_{v_1} & A^0_{v_2}
		\end{bmatrix}, \quad
		A^1_v = \begin{bmatrix}
			\pazocal{E} & A^1_{v_2} \\
			\pazocal{E} & \pazocal{E}
		\end{bmatrix},
	\]
	\[
		B^0_v = \begin{bmatrix}
			B^0_{v_1} & \pazocal{T} \\
			B^1_{v_1} & B^0_{v_2}
		\end{bmatrix},\quad 
		B^1_v = \begin{bmatrix}
			\pazocal{T} & B^1_{v_2} \\
			\pazocal{T} & \pazocal{T}
		\end{bmatrix}.
	\]
	To see the equivalence of~\eqref{eq:example_1} and~\eqref{eq:example_2}, observe that the second block of~\eqref{eq:example_2-1} reads $A_{v_1}^1 \otimes x(v^k) \oplus A_{v_2}^0 \otimes x(v^kv_1) \preceq x(v^kv_1) \preceq B_{v_1}^1 \stimes x(v^k) \splus B_{v_2}^0 \stimes x(v^kv_1)$.
	From this transformation, we can easily conclude that $\pazocal{S}$ is boundedly consistent under $v^\omega$ if and only if the LDI with characteristic matrices $A^0_v,A^1_v,B^0_v,B^1_v$ is boundedly consistent, and that all consistent $v$-periodic trajectories of $\pazocal{S}$ coincide with consistent 1-periodic trajectories of the LDI; hence,
	\[
		\solSLDI[\wA\wC] = \solPTEG[A^0_{\wA\wC}][A^1_{\wA\wC}][B^0_{\wA\wC}][B^1_{\wA\wC}] = \emptyset,
	\]\[
		\solSLDI[\wA\wB] = \solPTEG[A^0_{\wA\wB}][A^1_{\wA\wB}][B^0_{\wA\wB}][B^1_{\wA\wB}] = [3,3].
	\]

	It is worth noting that, although P-TEGs labeled $\wA$ and $\wB$ are not boundedly consistent, the SLDI under schedule $(\wA\wB)^\omega$ is.
	Thus, in general it is not possible to infer bounded consistency of an SLDI under a fixed schedule $w$ solely based on the analysis of each mode appearing in $w$.

	By generalizing the procedure shown in Example~\ref{ex:2}, we can derive the following proposition through some simple algebraic manipulations (to set up an equivalent LDI) and applying Theorem~\ref{th:bounded_consistency}.
\end{exmp}

\begin{prop}\label{pr:SLDI-P-TEGs}
	An SLDI $\pazocal{S}$ is boundedly consistent under schedule $w=v^\omega$ if and only if it admits a $v$-periodic trajectory.
    Moreover, set $\solSLDI$ coincides with $\solPIC[\MP_v][\MI_v][\MC_v]$, where
    \[
        \MP_v = \Maux_{|v|,1} \otimes^t \MP_{v_{|v|}},\quad
        \MI_v = \Maux_{1,|v|} \otimes^t \MI_{v_{|v|}},
        \] 
        \[
			\begin{array}{ll}
        \MC_v  = & \bigoplus_{r = 1}^{|v|-1} ( \Maux_{r,r+1} \otimes^t \MP_{v_r} \oplus \Maux_{r+1,r}\otimes^t \MI_{v_r} \oplus \Maux_{r,r} \otimes^t \MC_{v_r} ) \\
		& \oplus \Maux_{|v|,|v|}\otimes^t \MC_{v_{|v|}},
			\end{array}
    \] 
    where, for all $i,j\in\{1,\dots,|v|\}$, $\Maux_{i,j}\in\Rmax^{|v|\times |v|}$, with $(\Maux_{i,j})_{hk} = 0$ if $h=i$ and $k=j$, $(\Maux_{i,j})_{hk} =-\infty$ else, for all $r\in\{1,\dots,|v|\}$, $\MP_{v_r} = B^{1\sharp}_{v_r}$, $\MI_{v_r} = A^1_{v_r}$, and $\MC_{v_r} = A^0_{v_r} \oplus B^{0\sharp}_{v_r}$.
\end{prop}

For instance, when $|v|=5$, matrix $\lambda \MP_v \oplus \lambda^{-1} \MI_v \oplus \MC_v$ has the form
\begin{equation}\label{eq:sparse_matrix}
	\begin{bmatrix}
		\MC_{v_1} & \MP_{v_1} & \pazocal{E} & \pazocal{E} & \lambda^{-1}\MI_{v_5} \\
		\MI_{v_1} & \MC_{v_2} & \MP_{v_2} & \pazocal{E} & \pazocal{E} \\
		\pazocal{E} & \MI_{v_2} & \MC_{v_3} & \MP_{v_3} & \pazocal{E} \\
		\pazocal{E} & \pazocal{E} & \MI_{v_3} & \MC_{v_3}& \MP_{v_4} \\
		\lambda \MP_{v_5} & \pazocal{E} & \pazocal{E} & \MI_{v_4} & \MC_{v_5}
	\end{bmatrix},
\end{equation} 
which can be rewritten, using the tensor product, as
\[
	\Maux_{1,1}\otimes^t\MC_{v_1} \oplus \Maux_{1,2}\otimes^t\MP_{v_1} \oplus \lambda^{-1} \Maux_{1,5}\otimes^t \MI_{v_5} \oplus \Maux_{2,1}\otimes^t\MI_{v_1}\oplus \ldots
\]

Proposition~\ref{pr:SLDI-P-TEGs} directly provides an algorithm to compute the minimum and maximum cycle times of an SLDI under a fixed periodic schedule.
Indeed, these values come from solving the NCP for parametric precedence graph $\graph(\lambda\MP_v\oplus\lambda^{-1}\MI_v\oplus\MC_v)$.
However, this approach results in a slow (although strongly polynomial time) algorithm when the length of subschedule $v$ is large; indeed, its time complexity is $\pazocal{O}((|v|n)^4) = \pazocal{O}(|v|^4n^4)$, as the considered precedence graph has $|v|n$ nodes.
In the next subsection, we show how to exploit the sparsity of $\lambda\MP_v\oplus\lambda^{-1}\MI_v\oplus\MC_v$, illustrated for $|v|=5$ in~\eqref{eq:sparse_matrix}, to develop an algorithm of linear complexity in the subschedule length.

\subsection{Improved algorithm}\label{su:improved_algorithm}

\begin{figure}
	\centering
	\resizebox{.45\linewidth}{!}{
	\input{figures/switching_graph.tex}
	}
	\caption{Lumped-node representation of $G_v$ when $|v|=5$. Labels colored in \textcolor{myblue}{blue}, \textcolor{myred}{red}, and black correspond to arcs whose weight depends proportionally, depends inversely, and does not depend on $\lambda$, respectively.}
	\label{fig:switching_graph}
\end{figure}
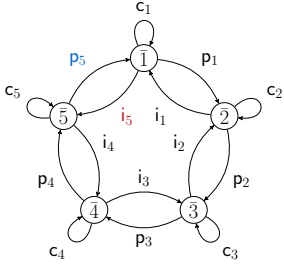

Let us start by defining the multi-precedence graph $G_v$ associated with parametric precedence graph $\graph(\lambda\MP_v\oplus\lambda^{-1}\MI_v\oplus \MC_v)$: $G_v=\graph(\Maux_{1,1}\otimes^t\MC_{v_1}, \Maux_{1,2}\otimes^t\MP_{v_1}, \Maux_{1,|v|}\otimes^t \lambda^{-1}\MI_{v_{|v|}}, \Maux_{2,1}\otimes^t\MI_{v_1},\ldots) = (\nodes,\bar{\Sigma},\ltm,\arcs)$ is such that $\nodes=\{1,\ldots,|v|n\}$, $\bar{\Sigma}=\{\wP_1,\ldots,\wP_{|v|},\wI_1,\ldots,\wI_{|v|},\wC_1,\ldots,\wC_{|v|}\}$, for all $r\in\{1,\ldots,|v|-1\}$, $\ltm(\wP_r) = \Maux_{r,r+1}\otimes^t\MP_{v_r},\ \ltm(\wI_r) = \Maux_{r+1,r}\otimes^t\MI_{v_r},\ \ltm(\wC_r) = \Maux_{r,r}\otimes^t\MC_{v_r}$, 
$\ltm(\wP_{|v|}) = \lambda \Maux_{|v|,1}\otimes^t\MP_{v_{|v|}}$, $\ltm(\wI_{|v|}) = \lambda^{-1}\Maux_{1,|v|}\otimes^t\MI_{v_{|v|}}$, $\ltm(\wC_{|v|}) = \Maux_{|v|,|v|}\otimes^t\MC_{v_{|v|}}$.
The multi-precedence graph $G_v$ is schematized by the lumped-node representation of Figure~\ref{fig:switching_graph} in the case $|v|=5$.
In this representation, $\bar{j}$ indicates the set of nodes $\{(j-1)n + 1,\ldots,jn\}$ of $G_v$, and an arc from $\bar{i}$ to $\bar{j}$ with label $\wZ$ indicates that, in $G_v$, every arc from a node in $\bar{i}$ to a node in $\bar{j}$ is labeled $\wZ$.

Let $\lang_1$ be the (regular) language containing the labels of all circuits in $G_v$ from any node in $\bar{1}=\{1,\ldots,n\}$.
With the visual aid of the lumped-node representation, we can use automata-theory techniques to determine an expression for $\lang_1$:
reinterpret the lumped-node representation of $G_v$ as a deterministic finite automaton with states $\bar{j}$ for all $j \in \{1,\ldots,|v|\}$; then, $\lang_1$ is the language recognized by the automaton when $\bar{1}$ is both initial and final state.
Once we get $\lang_1$, values of $\lambda$ such that $\tr(\ltm(\lang_1))=0$ will correspond to those for which there are no circuits in $G_v$, visiting at least one node from $\bar{1}$, with positive weight; we will see later how to derive from this observation a low-complexity algorithm that finds all $\lambda$'s such that $G_v\in\nonegsetm$ -- clearly, due to Proposition~\ref{pr:multi_precedence} these values correspond to the cycle times that we are looking for.
Language $\lang_1$ can be written as follows: $\lang_1 = (\mathbb{P}_{|v|} \cup \mathbb{I}_{|v|}\cup\mathbb{C}^P_{|v|}\cup\mathbb{C}^I_{|v|}\cup\wC_1)^*$, where
\begin{itemize}
	\item[-] $\mathbb{P}_{|v|}=\wC_1^*\lang\wC_1^*$ is such that $\lang\subseteq\bar{\Sigma}^*$ contains all strings $s\in\lang_1$ with $s_1=\wP_1$, $s_{|s|}=\wP_{|v|}$, $|s|_{\wP_{1}}=|s|_{\wP_{|v|}}=1$;
	\item[-] $\mathbb{I}_{|v|}=\wC_1^*\lang\wC_1^*$ is such that $\lang\subseteq\bar{\Sigma}^*$ contains all strings $s\in\lang_1$ with $s_1=\wI_{|v|}$, $s_{|s|}=\wI_{1}$, $|s|_{\wI_{1}}=|s|_{\wI_{|v|}}=1$;
	\item[-] $\mathbb{C}_{|v|}^P=\wC_1^*\lang\wC_1^*$ is such that $\lang\subseteq\bar{\Sigma}^*$ contains all strings $s\in\lang_1$ with $s_1=\wP_1$, $s_{|s|}=\wI_1$, $|s|_{\wP_{|v|}}=|s|_{\wI_{|v|}}=0$, and $|s|_{\wP_1}=|s|_{\wI_1}=1$;
	\item[-] $\mathbb{C}_{|v|}^I=\wC_1^*\lang\wC_1^*$ is such that $\lang\subseteq\bar{\Sigma}^*$ contains all strings $s\in\lang_1$ with $s_1=\wI_{|v|}$, $s_{|s|}=\wP_{|v|}$, $|s|_{\wP_1}=|s|_{\wI_1}=0$, and $|s|_{\wP_{|v|}}=|s|_{\wI_{|v|}}=1$.
\end{itemize}

To get an expression for $\lang_1$, it is convenient to first define some auxiliary languages: for all $r\in\{1,\ldots,|v|\}$, $\pazocal{P}_r = \wC_r^*\wP_r\wC_{r+1}^*$, $\pazocal{I}_r = \wC_{r+1}^*\wI_r\wC_r^*$ with 
$\wC_{|v|+1}\coloneqq \wC_1$.
Then, $\mathbb{P}_1 = \paP_{|v|}$, $\mathbb{I}_1 = \paI_1$, $\mathbb{C}^P_1=\mathbb{C}^I_1=\{\emptystr\}$, and for all $r\in\{1,\ldots,|v|-1\}$,
\[
	\begin{array}{rclcrcl}
	\mathbb{P}_{r+1} &=& \paP_{|v|-r}\mathbb{C}^P_r\mathbb{P}_r, && \mathbb{I}_{r+1} &=& \paI_{r+1}\mathbb{C}^I_r\mathbb{I}_r,\\
	\mathbb{C}^P_{r+1} &=& (\paP_{|v|-r}\mathbb{C}^P_r\paI_{|v|-r})^*, && \mathbb{C}^I_{r+1} &=& (\paI_{r+1}\mathbb{C}^I_r\paP_{r+1})^*.
	\end{array}
\]
For instance, in the case $|v|=5$, we get
\[
	\begin{array}{rl}
	\mathbb{P}_5 = & \paP_1(\paP_2(\paP_3(\paP_4\paI_4)^*\paI_3)^*\paI_2)^*\\ & \paP_2(\paP_3(\paP_4\paI_4)^*\paI_3)^*\paP_3(\paP_4\paI_4)^*\paP_4\paP_5,\\
	\mathbb{I}_5 = & \paI_5(\paI_4(\paI_3(\paI_2\paP_2)^*\paP_3)^*\paP_4)^*\\ & \paI_4(\paI_3(\paI_2\paP_2)^*\paP_3)^*\paI_3(\paI_2\paP_2)^*\paI_2\paI_1,\\
	\mathbb{C}^P_{5} = & \paP_1(\paP_2(\paP_3(\paP_4\paI_4)^*\paI_3)^*\paI_2)^*\paI_1,\\
	\mathbb{C}^I_5 = & \paI_5(\paI_4(\paI_3(\paI_2\paP_2)^*\paP_3)^*\paP_4)^*\paP_5.\\
	\end{array}
\]

Observe that, for all $A,B,C\in\Rmax^{n\times n}$ and $i,j\in\{1,\ldots,|v|\}$, the following properties hold: from $\Maux_{i,i}^2 = \Maux_{i,i}$, Proposition~\ref{pr:tensor_mixed}, and the fact that in the max-plus algebra the tensor product distributes over infinite sums,
\[
	\begin{array}{ll}
	(\Maux_{i,i}\otimes^t A)^* &=  \bigoplus_{k=0}^{\infty} (\Maux_{i,i}\otimes^t A)^k = \bigoplus_{k=0}^{\infty} \Maux_{i,i}^k\otimes^t A^k \\ &= \Maux_{i,i} \otimes^t \bigoplus_{k=0}^{\infty} A^k = \Maux_{i,i}\otimes^t A^*;
	\end{array}
\]
moreover, since $\Maux_{i,j}\otimes \Maux_{j,j}\otimes \Maux_{j,i} = \Maux_{i,i}$,
\[
	\begin{array}{ll}
	((\Maux_{i,j}\otimes^t A) (\Maux_{j,j}\otimes^t B) (\Maux_{j,i}\otimes^t C))^* &= \Maux_{i,i} \otimes^t (ABC)^*.
	\end{array}
\]
Hence, for some matrices $L_P,L_I,L_{CP},L_{CI}\in\Rbar^{n\times n}$, $\ltm(\mathbb{P}_{|v|}) = \Maux_{1,1}\otimes^t \lambda L_P$, $\ltm(\mathbb{I}_{|v|}) = \Maux_{1,1}\otimes^t \lambda^{-1}L_I$, $\ltm(\mathbb{C}^P_{|v|}) = \Maux_{1,1}\otimes^t L_{CP}$, $\ltm(\mathbb{C}^I_{|v|}) = \Maux_{1,1}\otimes^t L_{CI}$, and 
\[
	\ltm(\lang_1) = \Maux_{1,1} \otimes^t (\lambda L_P \oplus \lambda^{-1}L_I \oplus L_{CP} \oplus L_{CI} \oplus C_{v_1})^*.
\]
Finally, from Proposition~\ref{pr:trace_tensor} and $\tr(\Maux_{1,1}) = 0$, we get $\tr(\ltm(\lang_1)) = \tr((\lambda L_P \oplus \lambda^{-1}L_I \oplus L_{CP} \oplus L_{CI} \oplus C_{v_1})^*) = 0$ if and only if $\graph(\lambda L_P \oplus \lambda^{-1}L_I \oplus L_{CP} \oplus L_{CI} \oplus C_{v_1})\in\nonegset$.
Observe that we obtained an NCP that can be solved in $\pazocal{O}(n^4)$ using Algorithm~\ref{al:PIC-NCP}.

To find all $\lambda$'s for which $G_v\in\nonegsetm$, we still need to verify that there are no circuits with positive weight among those visiting only nodes that are not in $\bar{1}$ (if this is not true, then $\solSLDI=\emptyset$).
This can be done by checking that, for all $r\in\{1,\ldots,|v|\}$, $\graph(\ltm(\wC_r))\in\nonegset$, and $\graph(\ltm(\mathbb{C}^P_r))\in\nonegset$ (or, equivalently, $\graph(\ltm(\mathbb{C}^I_r))\in\nonegset$).
Indeed, a circuit $\sigma$ starting from a node in $\bar{r}$ that does not visit any node in $\bar{1}$ either does not visit any node in other sets $\bar{j}\neq\bar{r}$, in which case the label $s$ of $\sigma$ belongs to $\wC_r^*$, or it does.
In the second case, take the smallest $j$ for which $\sigma$ visits nodes in $\bar{j}$, say $j'$.
If $j'=r$, then $s\in\mathbb{C}^P_r$; otherwise, there exists a circuit $\sigma'$ with label $s'$ that visits the same nodes of $\sigma$ using the same arcs, but starting from a node in $\bar{j}'$.
Thus, the weights of $\sigma$ and $\sigma'$ coincide, and $s'\in\mathbb{C}^P_{j'}$.
This proves that $\graph(\ltm(\wC_r))\in\nonegset$ and $\graph(\ltm(\mathbb{C}^P_r))\in\nonegset$ $\forall r\in\{1,\ldots,|v|\}$ is a necessary and sufficient condition for the non-positiveness of the weight of all circuits in $G_v$ not visiting nodes in $\bar{1}$.
Note that, similarly to $\ltm(\lang_1)$, $\ltm(\wC_r)$ and $\ltm(\mathbb{C}^P_r)$ can be written as the tensor product between $\Maux_{r,r}$ and an $n\times n$ matrix, and we can exploit this fact to decrease the complexity for checking the non-positiveness of the circuits of their associated precedence graphs. 

The discussed procedure to compute the minimum and maximum cycle times of an SLDI $\pazocal{S}$ under schedule $v^\omega$ is summarized in Algorithm~\ref{al:improved}.
Note that the time complexity to run lines 1--11 is $\pazocal{O}(|v|n^3)$, as the three for-loops perform operations of complexity $\pazocal{O}(n^3)$, namely, multiplying $n\times n$ matrices, computing their star, and checking whether the associated precedence graphs contain circuits with positive weight.
Line 12 requires $\pazocal{O}(n^4)$ operations; thus, the overall time complexity of Algorithm~\ref{al:improved} is $\pazocal{O}(|v|n^3+n^4)$, which is linear in the length of subschedule $v$, and its space complexity is $\pazocal{O}(|v|n^2)$.

\begin{algorithm2e}[t] \DontPrintSemicolon \label{al:improved} \small
\KwIn{$\MP_{\wZ},\MI_{\wZ},\MC_{\wZ}\in \Rmax^{n\times n}$ for all $\wZ\in\Sigma$, $v\in\Sigma^*$}
\KwOut{$\solSLDI$}
 \For{$r=1$ \KwTo $|v|$}{%
	\lIf{$\graph(\MC_{v_r})\notin\nonegset$}{%
		\Return $\emptyset$%
	}%
 }%
$v_{|v|+1}\leftarrow v_1$\\
 \For{$r=1$ \KwTo $|v|$}{
	$\MP_r \leftarrow \MC_{v_r}^* \MP_{v_r}\MC_{v_{r+1}}^*,\ \MI_{r} \leftarrow \MC_{v_{r+1}}^*\MI_{v_r}\MC_{v_r}^*$\\
 }
 $L_{\MC\MP}\leftarrow E_\otimes,\ L_{\MC\MI}\leftarrow E_{\otimes},\ L_{\MP}\leftarrow \MP_{v_{|v|}},\ L_{\MI}\leftarrow \MI_{v_1}$\\
 \For{$r=2$ \KwTo $|v|$}{
	\If{$\graph(\MP_{|v|-r+1}L_{\MC\MP}\MI_{|v|-r+1})\notin\nonegset$ \normalfont{or} $\graph(\MI_{r}L_{\MC\MI}\MP_{r})\notin\nonegset$}{
		\Return $\emptyset$\\
	}
	$L_{\MP} \leftarrow \MP_{|v|-r+1}L_{\MC\MP}L_{\MP}$, $L_{\MI}\leftarrow \MI_{r}L_{\MC\MI}L_{\MI}$\\
	$L_{\MC\MP} \leftarrow (\MP_{|v|-r+1}L_{\MC\MP}\MI_{|v|-r+1})^*$, $L_{\MC\MI} \leftarrow (\MI_{r}L_{\MC\MI}\MP_{r})^*$\\
 }
 \Return $\mathsf{Solve\_NCP}(L_{\MP},L_{\MI},L_{\MC\MP}\oplus L_{\MC\MI} \oplus \MC_{v_1})$\\
 \caption{Compute $\solSLDI$}
\end{algorithm2e}

\section{Practically-motivated example}\label{se:example}

The example we present is a multi-product processing network taken from~\cite{KATS20081196}.
Examples of such networks are electroplating lines and cluster tools.
Consider a manufacturing system consisting of 5 processing stations $S_1,\ldots,S_5$ and a robot of capacity one.
The system can treat two types of parts, part $\wA$, which requires to be processed in $S_1$, $S_3$, and $S_5$ in this order, and part $\wB$, which must follow route $S_2$, $S_1$, $S_4$, $S_5$.
The task of the robot is to transport parts of type $\wA$ and $\wB$ from an input storage $S_0$ to their first processing stations, between the processing stations (in the right order), and finally from the last processing station to an output storage $S_6$.
The time the robot takes to travel from $S_i$ to $S_j$ is $\tau_{ij}$ when it is not carrying any part, and $\tau_{ij}^\wZ$ when it is carrying part $\wZ\in\{\wA,\wB\}$.
Moreover, the processing time for part $\wZ$ in station $S_i$ must be within the interval $\iota_i^\wZ=[L_{i}^\wZ,R_{i}^\wZ]\subset \R_{\geq 0}$.

We suppose that initially station $S_3$ is processing a part of type $\wA$, and $S_2$, $S_4$ are processing parts of type $\wB$.  
We denote by $S_i\xrightarrow{\wZ} S_j$ robot operation "unload a part of type $\wZ$ from $S_i$, transport it to and load it into $S_j$" and by $\rightarrow S_j$ operation "travel from the current location to $S_j$ and wait if necessary".
A schedule for this process is an infinite sequence of modes $w\in\{\wA,\wB\}^\omega$, where mode $\wA$ represents the sequence of operations $\rightarrow S_3\xrightarrow{\wA} S_5 \rightarrow S_0 \xrightarrow{\wA} S_1 \rightarrow S_5 \xrightarrow{\wA} S_6 \rightarrow S_1 \xrightarrow{\wA} S_3$ and mode $\wB$ represents $\rightarrow S_4 \xrightarrow{\wB} S_5 \rightarrow S_2 \xrightarrow{\wB} S_1 \rightarrow S_5 \xrightarrow{\wB} S_6 \rightarrow S_0 \xrightarrow{\wB} S_2 \rightarrow S_1 \xrightarrow{\wB} S_4$.
Initially, the robot is positioned at $S_3$ if $w_1=\wA$ or at $S_4$ if $w_1=\wB$.
We consider the following parameters for the processing network: $\tau_{ij} = |i-j|$, $\tau_{ij}^\wA = \tau_{ij}+1$, $\tau_{ij}^\wB=\tau_{ij}+2$, $\iota_1^\wA=[10,15]$, $\iota_3^\wA=[40,140]$, $\iota_5^\wA=[20,30]$, $\iota_2^\wB=[50,150]$, $\iota_1^\wB=[10,20]$, $\iota_4^\wB=[30,150]$, $\iota_5^\wB=[20,30]$.

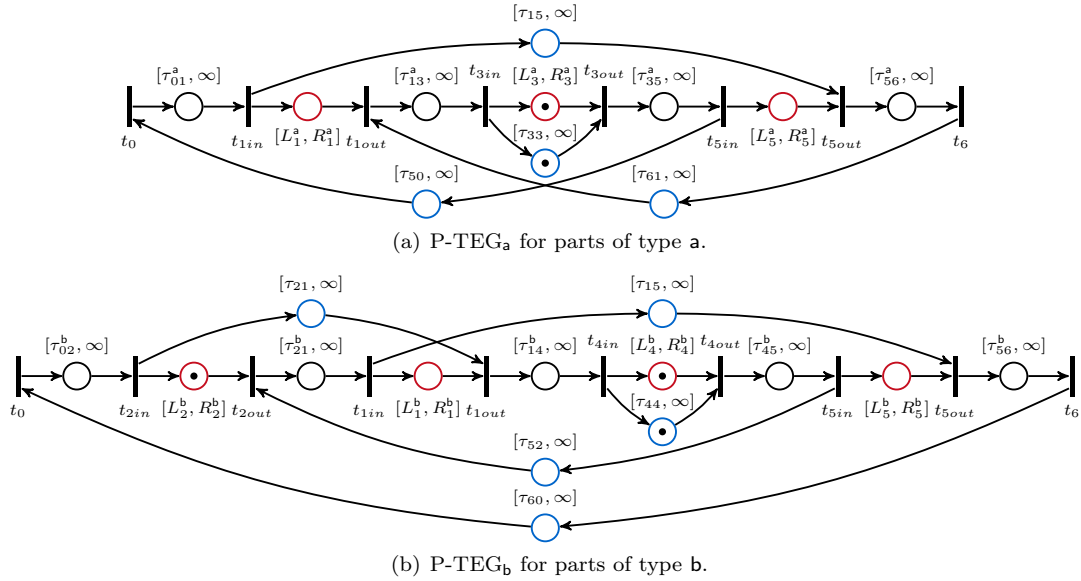
\begin{figure*}[ht]
	\centering
\subfigure[$\mbox{P-TEG}_\wA$ for parts of type $\wA$.]{
		\centering
		\resizebox{.64\textwidth}{!}{
		\input{figures/P-TEGs_processing_networka.tex}
		}
		\label{sub:1}}
	\subfigure[$\mbox{P-TEG}_\wB$ for parts of type $\wB$.]{
		\centering
		\resizebox{.8\textwidth}{!}{
		\input{figures/P-TEGs_processing_networkb.tex}
		}
		\label{sub:2}}
	\caption{P-TEGs modeling the processing network considering only one part-type.
	A token in a place colored \textcolor{myred}{red}, \textcolor{black}{black}, and \textcolor{myblue}{blue} represents a part being processed in a station, the robot moving with, and without carrying a part, respectively.}
	\label{fig:P-TEGs_processing_network}
\end{figure*}

Let us first model the processing network when only part $\wA$, respectively, $\wB$ is considered.
In this way, we obtain two P-TEGs, $\mbox{P-TEG}_\wA$ and $\mbox{P-TEG}_\wB$ (shown in Figure~\ref{fig:P-TEGs_processing_network}), each of which represents the behavior of the system when processing only parts of one type.
Using Algorithm~\ref{al:PIC-NCP}, we can find that the cycle times of the network when processing only parts of type $\wA$, $\wB$ are all values in $[73,\infty[$, and $[72,192]$, respectively.
Now, from the obtained P-TEGs, we can model the processing network in the case where both part-types are considered as an SLDI $\pazocal{S}=(\{\wA,\wB\},A^0,A^1,B^0,B^1)$.
To do so, we must define matrices $A^0_\wZ,A^1_\wZ,B^0_\wZ,B^1_\wZ\in\Rmax^{n\times n}$ for $\wZ\in\{\wA,\wB\}$ appropriately: we start by adding in $\mbox{P-TEG}_\wA$ (respectively, $\mbox{P-TEG}_\wB$) the missing transitions from $\mbox{P-TEG}_\wB$ (respectively, $\mbox{P-TEG}_\wA$) -- the obtained P-TEGs have both $n = 12$ transitions (in  general, $n=2+2\ \times$ number of processing stations).
For each new transition $t_i$ of $\mbox{P-TEG}_\wZ$, we define $(\MA^1_{\wZ})_{ii}=(\MB^1_{\wZ})_{ii} = 0$; this is done to store in auxiliary variables $x_i(w_{k]}\wZ)=x_i(w_{k]})$ the last entrance and exit times of parts in stations that are not used in mode $\wZ$.
Moreover, to model the transportation of the robot from $S_3$ to $S_4$ (respectively, from $S_4$ to $S_3$) after each switching of mode from $\wA$ to $\wB$ (respectively, from $\wB$ to $\wA$), we set $(\MA^1_\wA)_{4out,3in}=\tau_{34}$ (respectively, $(\MA^1_\wB)_{3out,4in}=\tau_{43}$).
The other elements of $\MA^0_\wZ,\MA^1_\wZ,\MB^0_\wZ,\MB^1_\wZ$ are taken from the characteristic matrices of $\mbox{P-TEG}_\wZ$, for $\wZ\in\{\wA,\wB\}$.
The modeling effort required to define $\pazocal{S}$ is repaid by the possibility to use Algorithm~\ref{al:improved} for computing the minimum and maximum cycle times corresponding to a schedule $w=v^\omega$.
For instance, we get $\solSLDI[\wA\wB] = [77,192]$.
This means that, using schedule $(\wA\wB)^\omega$, we can obtain one final product of each type every at least $77$ and at most $192$ time units.

\begin{figure}
\centering
	\resizebox{\linewidth}{!}{
    \input{figures/comparison.tex}
    }
	\caption{Time to compute $\solSLDI$ for increasing values of $|v|$ using different methods.}
	\label{fig:comparison}
\end{figure}
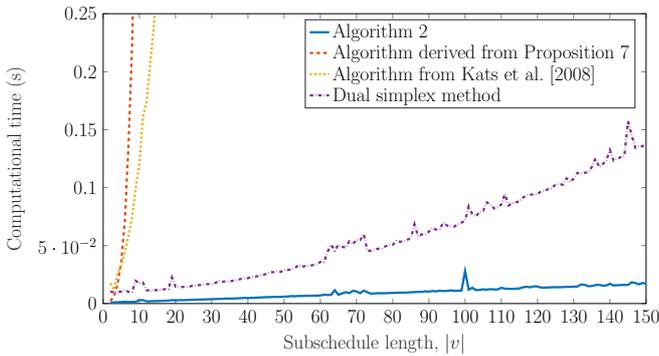

To appreciate the advantage of using Algorithm~\ref{al:improved}, in Figure~\ref{fig:comparison} we show the computational time to get $\solSLDI$ with increasing subschedule length $|v|$, using different methods: Algorithm~\ref{al:improved}, the algorithm derived from Proposition~\ref{pr:SLDI-P-TEGs} directly, the algorithm developed in~\cite{KATS20081196}, and a linear programming solver.
The first three algorithms were implemented on Matlab R2019a, for solving the linear programs we used CPLEX's dual simplex method; the tests were executed on a PC with an Intel i7 processor at 2.20Ghz.
From the results, we can see that the most time-consuming approach is the one using Proposition~\ref{pr:SLDI-P-TEGs} directly, while Algorithm~\ref{al:improved} achieves the fastest computation.
This shows how critical the exploitation of the sparsity of matrix $\lambda\MP_v\oplus\lambda^{-1}\MI_v\oplus\MC_v$ is for decreasing computation time. 

\section{Final remarks}\label{se:conclusions}

We have shown that SLDIs can model plants such as multi-product processing networks, and provided an inexpensive method to compute minimum and maximum cycle times when they follow a fixed and periodic schedule.
We remark that the complexity reduction achieved by exploiting the sparsity of matrix $\lambda\MP_v\oplus\lambda^{-1}\MI_v\oplus\MC_v$ through techniques from automata theory could be generalized to solve NCPs on matrices with different distributions of non-$\varepsilon$ elements; practical applications in a variety of scheduling problems are expected.
Regarding SLDIs, plenty of problems of theoretical and practical relevance remain open, such as the complexity of verifying the existence of a schedule $w$ under which the SLDI is boundedly consistent.
Finally, we argue that, as implicit switching max-plus linear systems generalize the dynamics of max-plus automata (cf.~\cite{4267666}), SLDIs generalize the dynamics of interval weighted automata.
This would imply that SLDIs can be used to represent and solve scheduling problems for systems modeled by safe P-time Petri nets (cf.~\cite{KOMENDA2020187}).

\bibliography{references}

\end{document}

%% file: math_commands.tex
\newcommand{\graph}{\pazocal{G}}
\newcommand{\R}{\mathbb{R}}
\newcommand{\nat}{\mathbb{N}}
\newcommand{\nato}{\mathbb{N}_0}
\newcommand{\Rmax}{{\R}_{\normalfont\fontsize{7pt}{11pt}\selectfont\mbox{max}}}
\newcommand{\Rmin}{{\R}_{\normalfont\fontsize{7pt}{11pt}\selectfont\mbox{min}}}

\newcommand{\Rbar}{\overline{\R}}
\newcommand{\emptystr}{\mathsf{e}}
\newcommand{\arcs}{E}
\newcommand{\nodes}{N}
\newcommand{\nonegset}{\Gamma}
\newcommand{\nonegsetm}{\Gamma_{M}}

\newcommand{\ltm}{\mu}
\newcommand{\lang}{\pazocal{L}}

\newcommand{\MP}{P}
\newcommand{\MI}{I}
\newcommand{\MC}{C}

\newcommand{\MA}{A}
\newcommand{\MB}{B}

\newcommandx{\solNCP}[1]{\Lambda_{\mbox{\normalfont\tiny NCP}}(#1)}
\newcommandx{\solPIC}[3][1=\MP,2=\MI,3=\MC]{\Lambda_{\mbox{\normalfont\tiny NCP}}(\lambda#1\oplus\lambda^{-1}#2\oplus#3)}
\newcommandx{\solPTEG}[4][1=A^0,2=A^1,3=B^0,4=B^1]{\Lambda_{\mbox{\normalfont\tiny P-TEG}}(#1,#2,#3,#4)}
\newcommandx{\solSLDI}[2][1=v,2=\pazocal{S}]{\Lambda^{#1}_{\mbox{\normalfont\tiny SLDI}}(#2)}
\newcommand{\wA}{\mathsf{\MakeLowercase{\MA}}}
\newcommand{\wB}{\mathsf{\MakeLowercase{\MB}}}
\newcommand{\wP}{\mathsf{\MakeLowercase{\MP}}}
\newcommand{\wI}{\mathsf{\MakeLowercase{\MI}}}
\newcommand{\wC}{\mathsf{\MakeLowercase{\MC}}}
\newcommand{\wZ}{\mathsf{z}}
\newcommand{\D}{\pazocal{D}}

\catcode`@=11
\def\plslash{\ifx\@currsize\normalsize
{\mathchoice
{\,\mbox{\raisebox{0.2ex}{$\scriptstyle\circ$}\kern-1ex$\setminus$}}
{\,\mbox{\raisebox{0.2ex}{$\scriptstyle\circ$}\kern-1ex$\setminus$}}%
{\,\mbox{\raisebox{0.14ex}{$\scriptscriptstyle\circ$}\kern-0.8ex%
${\scriptstyle\setminus}$}}%
{\,\mbox{\raisebox{0.14ex}{$\scriptscriptstyle\circ$}\kern-0.8ex%
${\scriptstyle\setminus}$}}}%
\else\ifx\@currsize\large\,\mbox{\raisebox{0.2ex}{$\scriptstyle\circ$}\kern-1ex$\setminus$}
\else\ifx\@currsize\small\,\mbox{\raisebox{0.2ex}{$\scriptstyle\circ$}\kern-1ex$\setminus$}
\else\,\mbox{\raisebox{0.2ex}{$\scriptstyle\circ$}\kern-0.1ex$\setminus$}
\fi\fi\fi}

\newcommandx{\TP}[1][1=d]{T_{#1}^{\MP}}
\newcommandx{\TI}[1][1=d]{T_{#1}^{\MI}}
\newcommandx{\TC}[1][1=d]{T_{#1}^{\MC}}
\newcommandx{\T}[2][2=d]{T_{#2}^{X_{#1}}}

\def\prslash{\ifx\@currsize\normalsize
{\mathchoice
{\mbox{\raisebox{0.2ex}{$\scriptstyle\circ$}\kern-1ex$/$}}
{\mbox{\raisebox{0.2ex}{$\scriptstyle\circ$}\kern-1ex$/$}}%
{\mbox{\raisebox{0.14ex}{$\scriptscriptstyle\circ$}\kern-0.8ex%
${\scriptstyle/}$}}%
{\mbox{\raisebox{0.14ex}{$\scriptscriptstyle\circ$}\kern-0.8ex%
${\scriptstyle/}$}}}%
\else\ifx\@currsize\large\mbox{\raisebox{0.2ex}{$\scriptstyle\circ$}\kern-1ex$/$}
\else\ifx\@currsize\small\mbox{\raisebox{0.2ex}{$\scriptstyle\circ$}\kern-1ex$/$}
\else\mbox{\raisebox{0.2ex}{$\scriptstyle\circ$}\kern-1ex$/$} \fi\fi\fi}

\def\plslashblack{\ifx\@currsize\normalsize
{\mathchoice
{\,\mbox{\raisebox{0.2ex}{$\scriptstyle\bullet$}\kern-1ex$\setminus$}}
{\,\mbox{\raisebox{0.2ex}{$\scriptstyle\bullet$}\kern-1ex$\setminus$}}%
{\,\mbox{\raisebox{0.14ex}{$\scriptscriptstyle\bullet$}\kern-0.8ex%
${\scriptstyle\setminus}$}}%
{\,\mbox{\raisebox{0.14ex}{$\scriptscriptstyle\bullet$}\kern-0.8ex%
${\scriptstyle\setminus}$}}}%
\else\ifx\@currsize\large\,\mbox{\raisebox{0.2ex}{$\scriptstyle\bullet$}\kern-1ex$\setminus$}
\else\ifx\@currsize\small\,\mbox{\raisebox{0.2ex}{$\scriptstyle\bullet$}\kern-1ex$\setminus$}
\else\,\mbox{\raisebox{0.2ex}{$\scriptstyle\bullet$}\kern-0.1ex$\setminus$}
\fi\fi\fi}

\def\prslash{\ifx\@currsize\normalsize
{\mathchoice
{\mbox{\raisebox{0.2ex}{$\scriptstyle\bullet$}\kern-1ex$/$}}
{\mbox{\raisebox{0.2ex}{$\scriptstyle\bullet$}\kern-1ex$/$}}%
{\mbox{\raisebox{0.14ex}{$\scriptscriptstyle\bullet$}\kern-0.8ex%
${\scriptstyle/}$}}%
{\mbox{\raisebox{0.14ex}{$\scriptscriptstyle\bullet$}\kern-0.8ex%
${\scriptstyle/}$}}}%
\else\ifx\@currsize\large\mbox{\raisebox{0.2ex}{$\scriptstyle\bullet$}\kern-1ex$/$}
\else\ifx\@currsize\small\mbox{\raisebox{0.2ex}{$\scriptstyle\bullet$}\kern-1ex$/$}
\else\mbox{\raisebox{0.2ex}{$\scriptstyle\bullet$}\kern-1ex$/$} \fi\fi\fi}
\catcode`@=12

\newcommand{\places}{\pazocal{P}}
\newcommand{\transitions}{\pazocal{T}}

\newcommand{\paP}{\pazocal{P}}
\newcommand{\paI}{\pazocal{I}}

\renewcommand{\qed}{\hfill\ensuremath{\blacksquare}}

\newcommand{\Maux}{Y}

\DeclareMathOperator{\tr}{tr}

\makeatletter
\newcommand{\splus}{%
  \DOTSB\mathop{\mathpalette\mattos@splus\relax}\slimits@
}
\newcommand\mattos@splus[2]{%
  \vcenter{\hbox{%
    \sbox\z@{$#1\oplus$}%
    \resizebox{!}{0.9\dimexpr\ht\z@+\dp\z@}{\raisebox{\depth}{$\m@th#1\boxplus$}}%
  }}%
  \vphantom{\oplus}%
}
\makeatother

\makeatletter
\newcommand{\stimes}{%
  \DOTSB\mathop{\mathpalette\mattos@stimes\relax}\slimits@
}
\newcommand\mattos@stimes[2]{%
  \vcenter{\hbox{%
    \sbox\z@{$#1\otimes$}%
    \resizebox{!}{0.9\dimexpr\ht\z@+\dp\z@}{\raisebox{\depth}{$\m@th#1\boxtimes$}}%
  }}%
  \vphantom{\otimes}%
}
\makeatother

\makeatletter
\newcommand{\bigsplus}{%
  \DOTSB\mathop{\mathpalette\mattos@bigsplus\relax}\slimits@
}
\newcommand\mattos@bigsplus[2]{%
  \vcenter{\hbox{%
    \sbox\z@{$#1\sum$}%
    \resizebox{!}{0.9\dimexpr\ht\z@+\dp\z@}{\raisebox{\depth}{$\m@th#1\boxplus$}}%
  }}%
  \vphantom{\sum}%
}
\makeatother

%% file: figures/P-TEG_example.tex
\begin{tikzpicture}[node distance=.5cm and 1.5cm,>=stealth',bend angle=30,thick]
\footnotesize
\node[transitionV,label=below:{$t_1$}] (t1) {};
\node[place,right=of t1,label=above:{$[0,+\infty]$}] (p12) {};
\node[transitionV,right=of p12,label=below:{$t_2$}] (t2) {};
\node[place,tokens=1,above=of t1,label=above:{$[\alpha_\wZ,\alpha_\wZ]$}] (p11) {};
\node[place,tokens=1,above=of t2,label=above:{$[\beta_\wZ,\beta_\wZ]$}] (p22) {};

\draw (t1) edge[->] (p12);
\draw (p12) edge[->] (t2);
\draw (t1.90-15) edge[bend right,->] (p11);
\draw (p11) edge[bend right,->] (t1.90+15);
\draw (t2.90-15) edge[bend right,->] (p22);
\draw (p22) edge[bend right,->] (t2.90+15);

\end{tikzpicture}

%% file: figures/switching_graph.tex
\begin{tikzpicture}[node distance=1.4cm and 1.4cm,>=stealth',bend angle=45,double distance=.5mm,arc/.style={->,>=stealth'},place/.style={circle,thick,minimum size=8mm,draw}]
\def \radius {2.5cm}
\Large

\foreach \t in {1,...,5}
{
\node [place,
       label=center:{$\bar{\t}$}
       ] at ({-360/5*(\t-1)+90}:\radius) (t\t) {};
\draw [arc] (t\t) to [out=-360/5*\t+360/5+90+25,in=-360/5*\t+360/5+90-25,loop] node[auto] {\textcolor{black}{$\wC_{\t}$}} (t\t);
}
\draw [arc] (t1) to [bend left=30] node[auto] {\textcolor{black}{$\wP_{1}$}} (t2);
\draw [arc] (t2) to [bend left=30] node[auto] {\textcolor{black}{$\wI_1$}} (t1);
\draw [arc] (t2) to [bend left=30] node[auto] {\textcolor{black}{$\wP_2$}} (t3);
\draw [arc] (t3) to [bend left=30] node[auto] {\textcolor{black}{$\wI_2$}} (t2);
\draw [arc] (t3) to [bend left=30] node[auto] {\textcolor{black}{$\wP_3$}} (t4);
\draw [arc] (t4) to [bend left=30] node[auto] {\textcolor{black}{$\wI_3$}} (t3);
\draw [arc] (t4) to [bend left=30] node[auto] {\textcolor{black}{$\wP_4$}} (t5);
\draw [arc] (t5) to [bend left=30] node[auto] {\textcolor{black}{$\wI_4$}} (t4);
\draw [arc] (t5) to [bend left=30] node[auto] {$\textcolor{myblue}{\wP_5}$} (t1);
\draw [arc] (t1) to [bend left=30] node[auto] {$\textcolor{myred}{\wI_5}$} (t5);

\end{tikzpicture}

%% file: figures/P-TEGs_processing_networka.tex
\begin{tikzpicture}[node distance=.5cm and .6cm,>=stealth',bend angle=30,thick]
\scriptsize
\node[transitionV,label=below:{$t_0$}] (t0) {};
\node[place,right=of t0,label=above:{$[\tau_{01}^\wA,\infty]$}] (p01) {};
\node[transitionV,right=of p01,label=below:{$t_{1in}$}] (t1in) {};
\node[place,myred,right=of t1in,label=below:{$[L_1^\wA,R_1^\wA]$}] (p1) {};
\node[transitionV,right=of p1,label=below:{$t_{1out}$}] (t1out) {};
\node[place,right=of t1out,label=above:{$[\tau_{13}^\wA,\infty]$}] (p13) {};
\node[transitionV,right=of p13,label=above:{$t_{3in}$}] (t3in) {};
\node[place,myred,tokens=1,right=of t3in,label=above:{$[L_3^\wA,R_3^\wA]$}] (p3) {};
\node[transitionV,right=of p3,label=above:{$t_{3out}$}] (t3out) {};
\node[place,right=of t3out,label=above:{$[\tau_{35}^\wA,\infty]$}] (p35) {};
\node[transitionV,right=of p35,label=below:{$t_{5in}$}] (t5in) {};
\node[place,myred,right=of t5in,label=below:{$[L_5^\wA,R_5^\wA]$}] (p5) {};
\node[transitionV,right=of p5,label=below:{$t_{5out}$}] (t5out) {};
\node[place,right=of t5out,label=above:{$[\tau_{56}^\wA,\infty]$}] (p56) {};
\node[transitionV,right=of p56,label=below:{$t_6$}] (t6) {};

\node[place,myblue,tokens=1,below=.4cm of p3,label=above:{$[\tau_{33},\infty]$}] (p33r) {};
\node[place,myblue,below=1cm of p13,label=above:{$[\tau_{50},\infty]$}] (p50r) {};
\node[place,myblue,above=of p3,label=above:{$[\tau_{15},\infty]$}] (p15r) {};
\node[place,myblue,below=1cm of p35,label=above:{$[\tau_{61},\infty]$}] (p61r) {};

\draw (t0) edge[->] (p01);
\draw (p01) edge[->] (t1in);
\draw (t1in) edge[->] (p1);
\draw (p1) edge[->] (t1out);
\draw (t1out) edge[->] (p13);
\draw (p13) edge[->] (t3in);
\draw (t3in) edge[->] (p3);
\draw (p3) edge[->] (t3out);
\draw (t3out) edge[->] (p35);
\draw (p35) edge[->] (t5in);
\draw (t5in) edge[->] (p5);
\draw (p5) edge[->] (t5out);
\draw (t5out) edge[->] (p56);
\draw (p56) edge[->] (t6);

\draw (t3in.-75) edge[bend right=10,->] (p33r);
\draw (p33r) edge[bend right=10,->] (t3out.180+75);
\draw (t5in.180+75) edge[bend left=10,->] (p50r);
\draw (p50r) edge[bend left=10,->] (t0.-75);
\draw (t1in.75) edge[bend left=10,->] (p15r);
\draw (p15r) edge[bend left=10,->] (t5out.180-75);
\draw (t6.180+75) edge[bend left=10,->] (p61r);
\draw (p61r) edge[bend left=10,->] (t1out.-75);

\end{tikzpicture}

%% file: figures/P-TEGs_processing_networkb.tex
\begin{tikzpicture}[node distance=.5cm and .6cm,>=stealth',bend angle=30,thick]
\scriptsize
\node[transitionV,label=below:{$t_0$}] (t0) {};
\node[place,right=of t0,label=above:{$[\tau_{02}^\wB,\infty]$}] (p02) {};
\node[transitionV,right=of p02,label=below:{$t_{2in}$}] (t2in) {};
\node[place,myred,tokens=1,right=of t2in,label=below:{$[L_2^\wB,R_2^\wB]$}] (p2) {};
\node[transitionV,right=of p2,label=below:{$t_{2out}$}] (t2out) {};
\node[place,right=of t2out,label=above:{$[\tau_{21}^\wB,\infty]$}] (p21) {};
\node[transitionV,right=of p21,label=below:{$t_{1in}$}] (t1in) {};
\node[place,myred,right=of t1in,label=below:{$[L_1^\wB,R_1^\wB]$}] (p1) {};
\node[transitionV,right=of p1,label=below:{$t_{1out}$}] (t1out) {};
\node[place,right=of t1out,label=above:{$[\tau_{14}^\wB,\infty]$}] (p14) {};
\node[transitionV,right=of p14,label=above:{$t_{4in}$}] (t4in) {};
\node[place,myred,tokens=1,right=of t4in,label=above:{$[L_4^\wB,R_4^\wB]$}] (p4) {};
\node[transitionV,right=of p4,label=above:{$t_{4out}$}] (t4out) {};
\node[place,right=of t4out,label=above:{$[\tau_{45}^\wB,\infty]$}] (p45) {};
\node[transitionV,right=of p45,label=below:{$t_{5in}$}] (t5in) {};
\node[place,myred,right=of t5in,label=below:{$[L_5^\wB,R_5^\wB]$}] (p5) {};
\node[transitionV,right=of p5,label=below:{$t_{5out}$}] (t5out) {};
\node[place,right=of t5out,label=above:{$[\tau_{56}^\wB,\infty]$}] (p56) {};
\node[transitionV,right=of p56,label=below:{$t_6$}] (t6) {};

\node[place,myblue,tokens=1,below=.4cm of p4,label=above:{$[\tau_{44},\infty]$}] (p44r) {};
\node[place,myblue,below=1cm of p14,label=above:{$[\tau_{52},\infty]$}] (p52r) {};
\node[place,myblue,above=of p21,label=above:{$[\tau_{21},\infty]$}] (p21r) {};
\node[place,myblue,above=of p4,label=above:{$[\tau_{15},\infty]$}] (p15r) {};
\node[place,myblue,below=.4cm of p52r,label=above:{$[\tau_{60},\infty]$}] (p60r) {};

\draw (t0) edge[->] (p02);
\draw (p02) edge[->] (t2in);
\draw (t2in) edge[->] (p2);
\draw (p2) edge[->] (t2out);
\draw (t2out) edge[->] (p21);
\draw (p21) edge[->] (t1in);
\draw (t1in) edge[->] (p1);
\draw (p1) edge[->] (t1out);
\draw (t1out) edge[->] (p14);
\draw (p14) edge[->] (t4in);
\draw (t4in) edge[->] (p4);
\draw (p4) edge[->] (t4out);
\draw (t4out) edge[->] (p45);
\draw (p45) edge[->] (t5in);
\draw (t5in) edge[->] (p5);
\draw (p5) edge[->] (t5out);
\draw (t5out) edge[->] (p56);
\draw (p56) edge[->] (t6);

\draw (t4in.-75) edge[bend right=10,->] (p44r);
\draw (p44r) edge[bend right=10,->] (t4out.180+75);
\draw (t5in.180+75) edge[bend left=10,->] (p52r);
\draw (p52r) edge[bend left=10,->] (t2out.-75);
\draw (t1in.75) edge[bend left=10,->] (p15r);
\draw (p15r) edge[bend left=10,->] (t5out.180-75);
\draw (t6.180+75) edge[bend left=10,->] (p60r);
\draw (p60r) edge[bend left=10,->] (t0.-75);
\draw (t2in.75) edge[bend left=10,->] (p21r);
\draw (p21r) edge[bend left=10,->] (t1out.180-75);

\end{tikzpicture}

%% file: figures/comparison.tex
%
%
\definecolor{mycolor1}{rgb}{0.00000,0.44700,0.74100}%
\definecolor{mycolor2}{rgb}{0.85000,0.32500,0.09800}%
\definecolor{mycolor3}{rgb}{0.92900,0.69400,0.12500}%
\definecolor{mycolor4}{rgb}{0.49400,0.18400,0.55600}%
\begin{tikzpicture}
\Large
\begin{axis}[%
width=6.625in,
height=3.566in,
at={(1.111in,0.481in)},
scale only axis,
xmin=0,
xmax=150,
xlabel style={font=\color{white!15!black}},
xlabel={Subschedule length, $|v|$},
ymin=0,
ymax=0.25,
ylabel style={font=\color{white!15!black}},
ylabel={Computational time (s)},
axis background/.style={fill=white},
legend style={legend cell align=left, align=left, draw=white!15!black}
]
\addplot [color=mycolor1, line width=2.0pt]
  table[row sep=crcr]{%
2	0.0008897\\
3	0.0010339\\
4	0.0011884\\
5	0.0013006\\
6	0.0016142\\
7	0.0013131\\
8	0.0014411\\
9	0.0015456\\
10	0.0029524\\
11	0.0029703\\
12	0.0018997\\
13	0.001969\\
14	0.0020826\\
15	0.0021912\\
16	0.0022711\\
17	0.0024423\\
18	0.0024875\\
19	0.0029573\\
20	0.0027767\\
21	0.0029237\\
22	0.0029363\\
23	0.0030276\\
24	0.0031631\\
25	0.0032157\\
26	0.0033576\\
27	0.0034589\\
28	0.0035529\\
29	0.0036572\\
30	0.0038192\\
31	0.0039116\\
32	0.0040025\\
33	0.0040827\\
34	0.0043574\\
35	0.0043157\\
36	0.0043979\\
37	0.0045038\\
38	0.0047019\\
39	0.0048779\\
40	0.0048305\\
41	0.0049267\\
42	0.0050822\\
43	0.0052092\\
44	0.0053528\\
45	0.0054818\\
46	0.0055457\\
47	0.0057415\\
48	0.0056899\\
49	0.0057963\\
50	0.0059417\\
51	0.00608\\
52	0.0064344\\
53	0.0062568\\
54	0.0065823\\
55	0.0064662\\
56	0.0067017\\
57	0.0066984\\
58	0.0067935\\
59	0.006867\\
60	0.0069923\\
61	0.0076724\\
62	0.0073903\\
63	0.007577\\
64	0.0114861\\
65	0.0075628\\
66	0.0087188\\
67	0.0098755\\
68	0.0090004\\
69	0.0110075\\
70	0.0103952\\
71	0.0089497\\
72	0.0112393\\
73	0.0097623\\
74	0.0084861\\
75	0.0086965\\
76	0.0089998\\
77	0.0088277\\
78	0.0089503\\
79	0.0090956\\
80	0.0092016\\
81	0.0093053\\
82	0.0094548\\
83	0.009442\\
84	0.0095452\\
85	0.0097614\\
86	0.0098586\\
87	0.0100399\\
88	0.0100753\\
89	0.0105631\\
90	0.0103136\\
91	0.0108315\\
92	0.010532\\
93	0.0110687\\
94	0.0108014\\
95	0.0109473\\
96	0.0113816\\
97	0.0110942\\
98	0.0110786\\
99	0.0112144\\
100	0.0279409\\
101	0.0119313\\
102	0.0136349\\
103	0.011629\\
104	0.0117853\\
105	0.0122649\\
106	0.0119945\\
107	0.0124362\\
108	0.0121761\\
109	0.0121699\\
110	0.0136413\\
111	0.0128076\\
112	0.0127321\\
113	0.0126252\\
114	0.0128336\\
115	0.0130394\\
116	0.0140302\\
117	0.0144886\\
118	0.0140397\\
119	0.0148641\\
120	0.0147151\\
121	0.0137649\\
122	0.0138194\\
123	0.0138353\\
124	0.0140778\\
125	0.0141121\\
126	0.0141076\\
127	0.0141934\\
128	0.0144646\\
129	0.0143734\\
130	0.014651\\
131	0.0146313\\
132	0.0164814\\
133	0.0161455\\
134	0.0149945\\
135	0.0151372\\
136	0.0159256\\
137	0.0161681\\
138	0.0154592\\
139	0.0155913\\
140	0.0171086\\
141	0.0156605\\
142	0.0158172\\
143	0.0159779\\
144	0.0160866\\
145	0.0160841\\
146	0.0183367\\
147	0.0177641\\
148	0.0165684\\
149	0.0177766\\
150	0.0171296\\
};
\addlegendentry{Algorithm 2}

\addplot [color=mycolor2, dashed, line width=2.0pt]
  table[row sep=crcr]{%
2	0.0022789\\
3	0.0062364\\
4	0.016549\\
5	0.0361977\\
6	0.0714103\\
7	0.1367755\\
8	0.2234487\\
9	0.3598495\\
10	0.5406782\\
11	0.7833683\\
12	1.1255212\\
13	1.5247941\\
14	1.9818483\\
15	2.428095\\
16	3.0958891\\
17	3.9913599\\
18	4.8444846\\
19	6.0514098\\
20	7.3040063\\
};
\addlegendentry{Algorithm derived from Proposition 7}

\addplot [color=mycolor3, dotted, line width=2.0pt]
  table[row sep=crcr]{%
2	0.0175802\\
3	0.0116395\\
4	0.0256474\\
5	0.0328238\\
6	0.0442625\\
7	0.0581616\\
8	0.0750381\\
9	0.0995027\\
10	0.1207159\\
11	0.1615339\\
12	0.1734218\\
13	0.2071105\\
14	0.2392245\\
15	0.2831648\\
16	0.3194853\\
17	0.371209\\
18	0.4232561\\
19	0.5421741\\
20	0.6057317\\
};
\addlegendentry{Algorithm from~\cite{KATS20081196}}

\addplot [color=mycolor4, dashdotted, line width=2.0pt]
  table[row sep=crcr]{%
2	0.0103715\\
3	0.0097556\\
4	0.0103573\\
5	0.0102854\\
6	0.0117051\\
7	0.0100959\\
8	0.010265\\
9	0.0196046\\
10	0.0173412\\
11	0.0181052\\
12	0.0114593\\
13	0.0112974\\
14	0.0116888\\
15	0.0115352\\
16	0.0118731\\
17	0.0124325\\
18	0.0125063\\
19	0.0225813\\
20	0.0145811\\
21	0.0141913\\
22	0.0143268\\
23	0.0144763\\
24	0.0148079\\
25	0.0154584\\
26	0.0156725\\
27	0.0161337\\
28	0.0166779\\
29	0.0167874\\
30	0.0170672\\
31	0.0177623\\
32	0.0195457\\
33	0.0188188\\
34	0.0193241\\
35	0.0191412\\
36	0.0198603\\
37	0.0203271\\
38	0.0212005\\
39	0.0217578\\
40	0.0219376\\
41	0.0228523\\
42	0.023218\\
43	0.0238258\\
44	0.0244112\\
45	0.0256011\\
46	0.0270048\\
47	0.0270694\\
48	0.0275655\\
49	0.0272553\\
50	0.0289497\\
51	0.0289764\\
52	0.0295306\\
53	0.0307177\\
54	0.0332527\\
55	0.0318339\\
56	0.0324587\\
57	0.032637\\
58	0.0331984\\
59	0.0337512\\
60	0.0359445\\
61	0.0358394\\
62	0.0447309\\
63	0.0505841\\
64	0.0458558\\
65	0.0508364\\
66	0.0491124\\
67	0.0481667\\
68	0.0536559\\
69	0.0511943\\
70	0.0549651\\
71	0.0546389\\
72	0.0598966\\
73	0.0454859\\
74	0.0456034\\
75	0.0466667\\
76	0.0473089\\
77	0.0478108\\
78	0.0493924\\
79	0.0506053\\
80	0.0507027\\
81	0.0539431\\
82	0.0532778\\
83	0.0537011\\
84	0.0551238\\
85	0.0561702\\
86	0.0683758\\
87	0.0580951\\
88	0.0596526\\
89	0.0597697\\
90	0.06201\\
91	0.065092\\
92	0.0637921\\
93	0.0642035\\
94	0.070098\\
95	0.0672079\\
96	0.0667518\\
97	0.0663807\\
98	0.068576\\
99	0.0698121\\
100	0.0714223\\
101	0.0838559\\
102	0.0768368\\
103	0.0752847\\
104	0.0789896\\
105	0.0800127\\
106	0.0872557\\
107	0.0855312\\
108	0.0820531\\
109	0.0830083\\
110	0.0855266\\
111	0.0946275\\
112	0.0846424\\
113	0.0865448\\
114	0.0873637\\
115	0.0882272\\
116	0.0914846\\
117	0.0932014\\
118	0.0938979\\
119	0.0941965\\
120	0.094445\\
121	0.0969633\\
122	0.0975538\\
123	0.0987139\\
124	0.0995131\\
125	0.1027713\\
126	0.1029658\\
127	0.102468\\
128	0.1079933\\
129	0.1049867\\
130	0.1077842\\
131	0.1127845\\
132	0.1119274\\
133	0.112828\\
134	0.1132826\\
135	0.1177295\\
136	0.1253345\\
137	0.117957\\
138	0.1203334\\
139	0.123371\\
140	0.1324343\\
141	0.1232387\\
142	0.1248836\\
143	0.1263396\\
144	0.1276201\\
145	0.157456\\
146	0.145368\\
147	0.1344298\\
148	0.1357757\\
149	0.1355729\\
150	0.1386939\\
};
\addlegendentry{Dual simplex method}

\end{axis}

\end{tikzpicture}%